\documentclass[prx,superscriptaddress,twocolumn,aps,showpacs,amsmath,amssymb,floatfix,longbibliography]{revtex4-1}
\usepackage{comment}
\usepackage{bm}
\usepackage{hyperref}
\usepackage{graphics}
\usepackage{graphicx}
\usepackage{braket}
\usepackage{MnSymbol}
\usepackage{amssymb}
\hypersetup{colorlinks=true}
\usepackage{amsmath,empheq}
\usepackage{mathrsfs}
\usepackage{textgreek}
\usepackage{xcolor}
\usepackage{enumitem}
\usepackage{dcolumn}
\usepackage[normalem]{ulem}
\usepackage{amsfonts}
\hypersetup{colorlinks,linkcolor=blue,urlcolor=blue,citecolor=blue}

 \renewcommand{\v}[1]{\textbf{#1}}
 \newcommand{\tr}{\text{tr}}

\begin{document}

\title{Drag resistance mediated by quantum spin liquids}

\author{Raffaele Mazzilli}
\affiliation{Max-Planck-Institut f\"ur Festk\"orperforschung, 70569 Stuttgart, Germany}

\author{Alex Levchenko}
\affiliation{Department of Physics, University of Wisconsin-Madison, Madison, Wisconsin 53706, USA}

\author{Elio~J.~K\"onig}
\affiliation{Max-Planck-Institut f\"ur Festk\"orperforschung, 70569 Stuttgart, Germany}
\date{\today }

\begin{abstract} 
Recent advances in material synthesis made it possible to realize two-dimensional monolayers of candidate materials for a quantum spin liquid (QSL) such as $\alpha$-RuCl$_3$,1T-TaSe$_2$ and 1T-TaS$_2$.  
In this work, we propose an experimental setup that exploits nonlocal electrical probes to gain information on the transport properties of a gapless QSL. The proposed setup is a spinon-mediated drag experiment: a current is injected in one of the two layers and a voltage is measured on the second metallic film. The overall momentum transfer mechanism is a two-step process mediated by Kondo interaction between the local moments in the quantum spin liquid and the spins of the electrons. In the limit of {negligible} momentum relaxed {with}in the QSL layer, we calculate the drag relaxation rate for Kitaev, $\mathbb{Z}_2$, and U(1) QSLs using Aslamazov-Larkin diagrams. We find, however, that the case of {dominant} 
momentum relax{ation with}in the QSL {layer} is far more relevant and thus develop a model based on the Boltzmann kinetic equation to {describe} the proposed setup. Within this framework we calculate the low temperature scaling behavior of the drag resistivity, both for U(1) and $\mathbb{Z}_2$  QSLs with Fermi surfaces. In some regimes we find a crossover in the temperature scaling that is different between the $\mathbb{Z}_2$ and U(1) QSL both because of the non-Fermi liquid nature of the U(1) QSL, and because of the qualitatively different momentum relaxation mechanism within the QSL layer. Our findings suggest that parameters of the system can be tuned to make the spinon-mediated drag a significant fraction of the total transresistance. 
\end{abstract}
\maketitle


\section{Introduction} 

Quantum Spin Liquids (QSLs) \cite{SavaryBalents2016,ZhouNg2017,KnolleMoessner2019} have enjoyed increased interest in the past years. Theoretically, one major driving force is the fascination for topological quantum states of matter~\cite{Wen2019} with strong entanglement, which are mathematically equivalent to certain topological quantum error correction codes proposed in quantum information theory \cite{Kitaev2003,Kitaev2006}. What is equally important, are the advances in materials science
and synthesis. Amongst the QSL candidate materials which were most actively discussed in recent years, particular attention is devoted to exfoliable materials such as the Kitaev candidate \cite{KasaharaMatsuda2018,YokoiMatsuda2021,CzajkaOng2021,BruinTakagi2022} $\alpha$-RuCl$_3$ as well as the transition metal dichalcogenides \cite{KratochvilovaPark2017,LawLee2017,MuruyamaMatsuda2020,RibakKanigel2017} 1T-TaS$_2$ and 1T-TaSe$_2$.

The study of monolayers and few layers of these van-der-Waals QSL candidate materials benefits from a variety of advantages: first, in the monolayer limit, interlayer valence bonds can not form, thereby precluding the stabilization of topologically trivial interlayer valence bond solids; second, the principle of van-der-Waals-LEGO \cite{GeimGrigorieva2013} allows to combine the functionality of 2D QSL sheets with superconductors, semiconductors, and metals in a single device and, as an example, {$\alpha$}-RuCl$_3$ on graphene devices were reported in the literature \cite{MashhadiKern2019}. Moreover, the stacking of van-der-Waals materials also allows to artificially create 2D QSL materials, for example in twisted transition-metal-dichalcogenide heterostructures \cite{ZareMosadeq2021, KieseKennes2022}. Third, the functionality of van-der-Waals heterostructures also allows for unprecedented experimental techniques to probe QSL behavior. These include tunneling spectroscopy, as studied theoretically \cite{MrossSenthil2011,MorampudiWilczek2017,CarregaPrincipi2020,ChenLado2020,FeldmeierKnolle2020,KoenigJaeck2020,JiaZou2021} and experimentally \cite{RuanCrommie2021,ChenCrommie2022} in different setups. Moreover, anyonic interference experiments to detect the topological order in the QSL state were proposed by interfacing them with quantum Hall and superconducting states \cite{AasenAlicea2020,KishoniBerg2021}. 
It appears plausible that this wide variety of advantages may thus counteract the main drawback of thin films, namely that standard bulk probes, such as neutron scattering \cite{KratochvilovaPark2017,BanerjeeNagler2016,BanerjeeNagler2017}  experiments are inefficient in the mono-layer or few-layer limit.

\begin{figure}
\includegraphics[width =.45\textwidth]{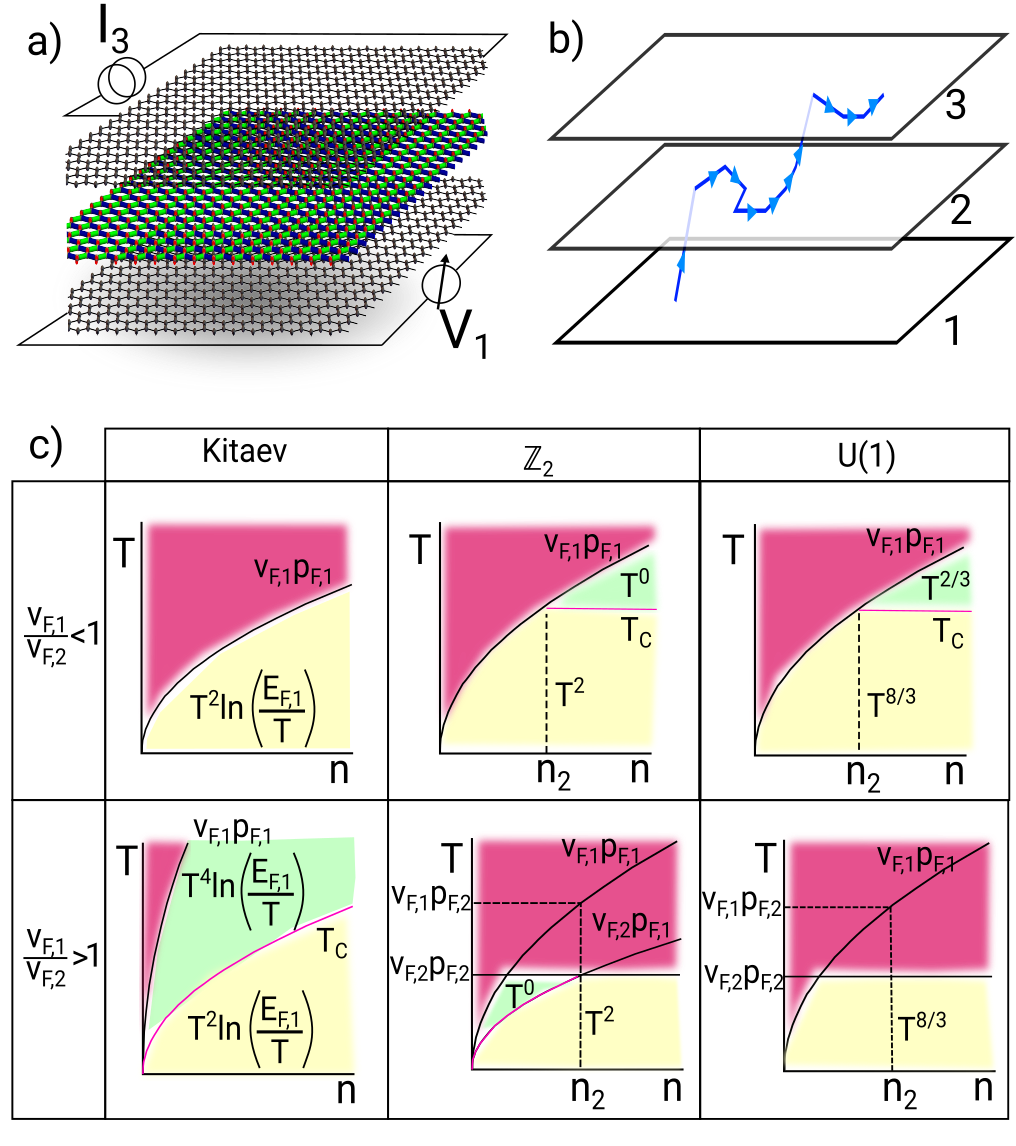}
 \caption{
a) Experimental setup under consideration: a QSL sheet is interposed between two metallic layers, a current is applied in the upper layer and a voltage is measured in the lower layer. b) Schematic representation of momentum transfer between the QSL layers and the metallic leads: momentum transfer is a two-step process.  
c) Summary of main results [see Eqs.~\eqref{eq:tauDKitaevNC}, \eqref{eq:tauDKitaevC}, \eqref{eq:tauDZ2BoltzNC}, \eqref{eq:tauDZ2BoltzC}, \eqref{eq:tauDU(1)NC}, \eqref{eq:tauDU(1)C} for details and discussion] i.e. the temperature dependence of drag in different regimes of experimentally tunable parameters, temperature $T$, and density $n$ in the metallic leads depicted for the three different types of QSLs under consideration.}
    \label{fig:Schematics}
\end{figure}

Motivated by the need for probing 2D QSL materials, in this paper we propose a \textit{spinon induced drag} experiment in which instead of a dielectric separating two metallic {films, as} in a Coulomb drag \cite{NarozhnhyLevchenko2016} experiment, we interpose a QSL material to mediate the interaction, see Fig.~\ref{fig:Schematics}a. 
In the limit when no momentum is deposited in the QSL layer, we 
calculate the temperature dependence of the nonlocal drag resistivity using standard Aslamazov-Larkin diagrams for three examples of gapless QSL states, including Kitaev QSLs and $\mathbb Z_2$ as well as U(1) QSLs with a spinon Fermi surface. Additionally, in the presence of a spinon Fermi surface, we also employ a Boltzmann-like framework, in which momentum relaxation inside the QSL layer is accounted for. Our study is thus applicable to the quantum states proposed for $\alpha$-RuCl$_3$ (Kitaev-like $\mathbb Z_2$ QSL with Dirac nodes) and the 1T-TaS$_2$ family (for which there are experimental signatures of U(1) QSL with spinon Fermi surface~\cite{RuanCrommie2021,ChenCrommie2022}).

The standard mechanism of momentum transfer between the layers mediated by Coulomb interaction results in a temperature $T$ dependent drag rate (defined in Sec.~\ref{sec:DragBasics}) \cite{NarozhnhyLevchenko2016}: 
$\tau^{-1}_D= \mathcal{C}(T^2/E_F)\mathrm{min}(1,{T_d}/{T})$, 
where $T_d=v_F/d$ and $E_F$, $v_F$ and $d$ are the Fermi energy, Fermi velocity and distance between the conducting layers (here and in what follows and we set $\hbar = 1 = k_B$). The dimensionless interaction parameter $\mathcal C \sim r_s^{-2} (k_F d)^{-4}$ (expressed in terms of $r_s \sim e^2/v_F \epsilon_r$ and Fermi wave-vector $k_F$) controls the magnitude of the effect. It is sensitive to the external screening and dielectric environment encoded in $\epsilon_r$. Drag due to exchange of bosons other than photons were considered in the literature before, including phonon-mediated mediated drag \cite{BonsagerMacDonald2000,Khveshchenko2000}. 
The most extensively studied example of drag resistance
in the case of non-Fermi liquids (nFLs) corresponds to the mutual friction between two copies of quantum Hall liquids at half filling. In this case Chern-Simons field theories of fermions coupled to the gauge field and also complimentary hydrodynamic ideas find their powerful applications \cite{Ussishkin1997,Sakhi1997,KimMillis1999,Patel2017}. 
Spinon contributions to the Coulomb drag of nFLs \cite{PereiraSela2010,ZouJongsoo2010} and within moir\'e bilayer systems \cite{ZhangVishwanath2020} were also considered, as well as a variety of (nonlocal) transport probes of spinons \cite{ChenXie2013,ChattarjeeSachdev2015,Okamoto2016,WermanBerg2018,AftergoodTakei2020,ZhuangMarston2021}. However, to the best of our knowledge, a spinon-mediated drag experiment as a probe of QSL physics has not been proposed before.

To conclude this introduction, we recall that in contrast to $\mathbb Z_2$ QSLs, the gauge field in U(1) QSLs is always strongly coupled to the spinons and large-$N$ methods are commonly used to access this non-trivial physics \cite{AltshulerMillis1994,KimLee1994,HermeleWen2004,Lee2009,MrossSenthil2010}. In the case of a spinon Fermi surface, a nFL phase arises which displays similar behavior as electrons at a nematic quantum critical point \cite{OganesyanFradkin2001,DellAnnaMetzner2007,MetlitskiSachdev2010,MaslovChubukov2011,HolderMetzner2015}, in particular a single particle lifetime $\tau \sim {\epsilon}^{-2/3}$ and a momentum relaxation time $\tau_2 \sim T^{-4/3}$, where {$\epsilon$} is the energy of the excitation {and $T$ the temperature}. At the same time, the non-trivial power-laws induced by gauge field fluctuations cancel in the polarization operator (i.e. the structure factor) in the low-energy long-wavelength limit~\cite{KimLee1994,AltshulerMillis1994,Lee2009,MrossSenthil2010,MetlitskiSachdev2010} such that we find little difference between the $\mathbb Z_2$ and U(1) QSL in the regime when no momentum is deposited into the QSL layer.

In the opposite limit, where momentum is lost in the QSL layer $\mathbb Z_2$ and $U(1)$ QSLs differ. In order to employ a Boltzmann-like approach to the drag resistivity across the U(1) QSL we follow a technique originally due to Prange and Kadanoff \cite{PrangeKadanoff1964,KimWen1995,NaveLee2007,KimPepin2009,HacklThomale2011} (see also Appendix \ref{app:KondoDerivation}), noting that the retarded spinon self energy $\Sigma^R(\v {k},{\epsilon})$ is independent of the modulus of the momentum $\v {k}$. Then the spectral weight $A(\v {k},{\epsilon})$ becomes a peaked function of $\vert \v  k \vert$ and can be used to define a pseudo-distribution function $f(\hat{k},{\epsilon})$. Within Keldysh technique, we derive the coupling of the Boltzmann-like equation describing the insulating QSL to the conventional Boltzmann equations describing the conducting layers.

This paper is organized as follows. In Sec. \ref{sec:Model and Experimental Setup} we describe the experimental setup of a spinon induced drag experiment, offer a phenomenological perspective and introduce the model for microscopic calculations. In Sec. \ref{sec:NoMomentumRelax} we use known results from Aslamazov-Larkin diagrammatics to calculate the drag relaxation rate for the Kitaev QSL, and for $\mathbb{Z}_2$ {as well as} U(1) QSL with Fermi surface. Sec. \ref{sec:Boltzmann} is devoted to the limit of momentum being relaxed inside the QSL and we present the derivation and solution of the system of coupled Boltzmann equations. We conclude with a discussion and outlook  and relegate important technical details to two appendices.


\section{Model and Experimental Setup}
\label{sec:Model and Experimental Setup}

\subsection{Spinon mediated drag}{\label{sec:SpinonInducedDrag}}
\label{sec:DragBasics}

The proposed setup is designed to measure spinon-mediated drag between two conducting two-dimensional electron systems. A layer of a QSL material is interposed between two metallic 
layers and a current {$\v {j}_3$} is passed through the active metallic layer (layer 3).
The spins of the free electrons Kondo couple with the local moments constituting 
the QSL material (layer 2), which then couple to the spins of the free electrons in the second metallic layer, also referred to as {``passive''} layer 1, thus inducing an electric field $\v E_1$. 
The proportionality constant between the field and current is the definition of drag conductivity
\begin{equation}
    \v {j}_3=\sigma_D \v {E}_1,
    \label{eq:definition_Sigma}
\end{equation}
and thereby to the drag rate 
\begin{equation}
    \frac{1}{\tau_D} \simeq \frac{n_3 e^2}{m_1}\frac{\sigma_D}{\sigma_1 \sigma_3}.
    \label{eq:1/tau_definizione}
\end{equation}
Here $n_i$ is the carrier density in layer $i=1,3$, $m_1$ is the electron's effective mass in layer 1, $e$ is the electric charge, and $\sigma_{1,3}$ the conductivities of the metallic layers. While our calculations are performed for metallic layers given by 2D electron gases (2DEG) with parabolic dispersion, the final temperature dependence of our results should also hold for doped graphene.

To gain insight into the problem, we briefly reiterate a simple Drude-like treatment based on the coupled equations of motion
\begin{align}
\dot{\v v}_1 &= - \frac{\v v_1}{\tau_1} - \frac{\v v_1 - \v v_{{2}}}{\tau_T} - \frac{e}{m_1} \v E_1, \\
\dot{\v v}_2 &=  - \frac{\v v_{{2}}}{\tau_{2}} - \frac{\v v_{{2}} - \v v_1}{\tau_T}- \frac{\v v_{{2}} - \v v_3}{\tau_T},\\
\dot{\v v}_3 &= - \frac{\v v_3}{\tau_3} - \frac{\v v_3 - \v v_{{2}}}{\tau_T}.
\end{align}
Here, the momentum relaxation rates within each layer are denoted $\tau_{1,2,3}$ and the rate of momentum transfer is $\tau_T$. For simplicity we assume parabolic bands in all layers. Here, and in most of the main text, we consider a symmetric setup with equivalent layers $1,3$, while details on non-symmetric setups are relegated to the appendices. The drag rate in the limit $\tau_1 = \tau_3 \ll \tau_T$ is thus
\begin{equation}
\frac{1}{\tau_D} = \frac{\tau_2}{\tau_T(\tau_T + 2  \tau_2)}. \label{eq:DrafFinalDrude}
\end{equation}

While this formula is phenomenologically valuable in the following discussion of limits, the remainder of the paper is devoted to the microscopic calculation of $1/\tau_D$.
The competition between the interlayer scattering and the intralayer scattering gives rise to two regimes. (i) For $\tau_T {\ll} \tau_2$ the momentum is quasi-instantaneously transferred from metallic layers to the QSL and from the latter to the second metal. The final result for $1/\tau_D$ is independent of $1/\tau_2$, since intralayer relaxation is unimportant there is no need for the description that makes use of the Boltzmann equation approach in the QSL layer. The microscopic theory for this regime is presented in Sec.~\ref{sec:NoMomentumRelax}. (ii) For $\tau_T \gg \tau_2$, the momentum transferred to the spinons relaxes before being transferred to the second lead. To account for these relaxation effects in the presence of a spinon Fermi surface, we employ a quantum Boltzmann equation, see Sec.~\ref{sec:Boltzmann}. We highlight that regime (i) and (ii) correspond to the limits $\tau_D \ll \tau_2$ and $\tau_D \gg \tau_2$, respectively.

\subsection{Models of QSLs}
\label{sec:QSLbasics}

In this section, we briefly summarize the models for the QSL state which we use. First, we consider QSL states with spinon Fermi surface.

As mentioned above a QSL comes in with the emergence of gauge fields. Theoretically the most prominent example of a QSL with spinon Fermi surface is a U(1) QSL, where the gauge degrees of freedom have their own Maxwellian dynamics and couple to the spin degrees of freedom 
\begin{equation}
    \hat{S}^a=\psi^{\dagger}_\sigma \sigma^a_{\sigma \sigma'} \psi_{\sigma'}. \label{eq:SUNSpins}
    \end{equation}
where $\psi_{\sigma}$ are fermionic spinons and the $\sigma^a$s describe the spin and we use Einstein summation convention. As we will shortly explain, we consider $\sigma^a$ to be SU(N) generators. The interacting action for such system is~\cite{LeeLee2005, NaveLee2007}
\begin{equation}
    S_f=\sum_{\sigma = 1}^N  \int_x \bar{\psi}_\sigma(x) \left[ i\partial_t +
    \frac{a_0}{\sqrt{N}} + \frac{1}{2m_f} \left(\boldsymbol{\nabla} + i \frac{\v a}{\sqrt{N}}\right)^2\right]\psi_\sigma(x),
    \label{eq:action_U(1)}
\end{equation}

Given the absence of an expansion control parameter as $\alpha=\frac{1}{137}$ in QED, we use a large $N$ expansion to deal with Eq.~\eqref{eq:action_U(1)}. Strictly speaking, a large $N$ expansion yields reliable results only when taken along with a small ${\varepsilon}$ expansion~\cite{Lee2009,MrossSenthil2010}, where ${\varepsilon}$ appears in the free action of the gauge boson
\begin{equation}
        S_{\rm MW}=\int d\textbf{k}\,d\omega \, \frac{|\textbf{k}|^{\varepsilon+1}}{e^2}|a(\textbf{k},\omega)|^2
        \label{eq:ActionGaugeU(1)},
\end{equation}
with the limits $\{\frac{1}{N},\varepsilon\} \rightarrow 0$ in a way that the product $N {\varepsilon}$ is constant. Despite this, following a common wisdom \cite{Lee2021} that the dynamics of the spinons is not affected by the double expansion, we use the large $N$ limit directly for ${\varepsilon}=0$.
    
As mentioned in the introduction, the U(1) QSL with Fermi surface forms a nFL. It is important to point out that the polarization operator {and thus the spin susceptibility} for the U(1) QSL for small $q$ and $\omega$ retains its low energy behavior due to mutual cancellation between vertex and self-energy corrections~\cite{KimLee1994,AltshulerMillis1994,Lee2009,MrossSenthil2010,MetlitskiSachdev2010}.
    
We compare the nFL of the U(1) QSL with Fermi surface to the case of $\mathbb Z_2$ QSLs with Fermi surface. In this case gauge fields are gapped and can be disregarded at lowest energies. The spinons form a Fermi liquid, and for simplicity of comparison to the U(1) case, we stick to the case with parabolic spinon dispersion and a spin representation as in Eq.~\eqref{eq:SUNSpins}. While the case of $\mathbb Z_2$ QSLs with Fermi surface is discussed much less in the literature, {it} does appear in certain exactly soluble models \cite{BaskaranShankar2009,HaoKivelson2009}.
    
Finally, we also calculate the drag relaxation time in a Kitaev spin liquid with no Fermi surface. {Despite the low-energy Majorana spinon excitations, the low-energy spin response is gapped \cite{KnolleMoessner2014} in the integrable Kitaev limit. However, taking into account integrability breaking Heisenberg and symmetric superexchange interactions} the {retarded} spin-spin correlation function is gapless and given by \cite{SongBalents2016}
  \begin{equation}
    C^R(\v {q},\omega)=C_K\sqrt{(v_{{F2}}q)^2-(\omega+i0^+)^2},
    \label{eq:DynStructKit}
\end{equation}
    where $C_K$ is a constant of dimension $1/(\text{energy}\times\text{length})^2$.
    
\subsection{Microscopic QSL-metal coupling}
\label{sec:Microscopic QSL-metal coupling}

The interlayer momentum and energy transfer {is microscopically encoded in} the Kondo coupling between the {local moments of the quantum magnet} 
and the spins of the free electrons 
\begin{equation}
    H = \frac{\mathcal{J}_K}{2N} \sum_{\v x}   \hat S^a(\v x) \cdot (c^{\dagger}_{\sigma_3}(\v x){\sigma}^a_{\sigma_3,\sigma_4}c_{\sigma_4}(\v x)),
    \label{eq:KondoCouplingLOCAL}
\end{equation}
where $c$ and $c^{\dagger}$ are the annihilation and creation operators for the electrons in the metal and $\mathcal{J}_K$ is the Kondo coupling constant. 

In the case of QSLs with a Fermi surface the effective coupling of spinons and electrons for the
continuum model is
\begin{equation}
    \begin{split}
    H=\frac{J_K}{2N} \int_{\v x}  (\psi^\dagger_{\sigma_1}(\v x)& {\sigma}^a_{\sigma_1,\sigma_2} \psi_{\sigma_2}(\v x)) (c^{\dagger}_{\sigma_3}(\v x){\sigma}^a_{\sigma_3,\sigma_4}c_{\sigma_4}(\v x)).
    \end{split}
    \label{KondoCouplingContinuum}
\end{equation}
Here, $J_K \sim \mathcal J_K a^2$, with $a$ being the lattice spacing and the subscripts $\sigma_i$ are spin indices and we use the large $N$ expansion for our calculations, so ${\sigma_i}=1,\dots ,N$ {and the generators of SU(N) have the property $\tr[\sigma^a \sigma^b] = 2\delta_{ab}$}.

\subsection{Momentum relaxation in QSLs with Fermi surface}
\label{sec:Momentum relaxation in QSLs with Fermi surface}

Before continuing with the microscopic calculation of the drag current, we summarize below the expected low-temperature behavior of the momentum relaxation rate $1/\tau_2$ in QSLs with spinon Fermi surface.

For the $\mathbb{Z}_2$ QSL with a Fermi surface, for small $\omega$ and $q$ the situation is similar to that of a Fermi liquid \cite{PalMaslov2012}. In a clean, interacting Fermi liquid on a lattice,
Umklapp scattering is the leading momentum relaxation channel. In the presence of impurities, the low temperature relaxation time is temperature independent, therefore:
 \begin{equation}
    \frac{1}{\tau_2}\propto
    \begin{cases}
    T^0,  &\text{Impurity-dominated scattering;}\\
    T^2, & \text{Umklapp-dominated scattering.}
    \end{cases}
    \label{eq:Z2Scattering rate}
\end{equation}

In contrast, for the nFL U(1) QSL with Fermi surface, the spinons interact with {gapless} gauge bosons. If the gauge bosons are assumed to form a thermal bath \cite{NaveLee2007}, which can serve as a sink of energy and momentum, then the relaxation happens via the momentum transferred from the spinons to the gauge bosons. Clearly, this requires that the gauge bosons are an open quantum system, which thermalizes faster than the rate of returning the energy and momentum to the spinons. If the overall system of spinons and gauge bosons is approximately closed, and thus momentum conserving, spinon-gauge boson interactions have to be supplemented by Umklapp and impurity scattering. It was shown \cite{Lee2021} that if Umklapp scattering is dominant, and there is weak impurity scattering, the momentum relaxation rate is qualitatively different with respect to the case of scattering with thermal gauge bosons. We summarize the two different situations described above, and additionally quote the result for impurity scattering:
\begin{equation}
    \frac{1}{\tau_2}\propto
    \begin{cases}
    T^0 & \text{Impurity scattering;}\\
    T^{4/3} &\text{Gauge boson scattering;}\\
    T  & \text{Umklapp scattering.}\\
    \end{cases}
\label{eq:U1ScatteringRate}
\end{equation}


\section{Theory without momentum relaxation in the QSL layer}
\label{sec:NoMomentumRelax}

In the case of $\tau_D^{-1}\gg \tau_2^{-1}$ with no momentum relaxation within the QSL layer the drag rate is calculated with the formula 
\cite{JauhoSmith1993,NarozhnhyLevchenko2016}
\begin{align}
 \frac{1}{\tau_{\rm{D}}}&=\frac{1}{ {p^2_{F1}}m_{{1}}}
\int_{0}^{\infty}\frac{d\omega/T}{\sinh^2(\omega/2T)}\nonumber \\
&\times\int_{0}^{\infty}dq\,q^3 |C(\v {q},\omega)|^2{\rm{Im}}[\Pi_{{1}}(\v  q ,\omega)]^2. 
\label{eq:ALDrag}    
\end{align}
For simplicity, we concentrate on the case of metallic layers with equal scattering rate, mass and Fermi energy, and polarization operator $\Pi_1 (\v  q, \omega) = \Pi_3 (\v  q, \omega)$.

For the standard Coulomb drag, $C(\v {q},\omega)$ can be approximated by the electrostatic potential between a
point charge on the active layer and a point charge of the passive layer. In the present case it is proportional to  
the spin-spin correlation function with additional factor $J_K^2/N$.

The support for the integration in Eq.~\eqref{eq:ALDrag} is given by domain of the imaginary parts of the polarization operator of metallic layers. As we explain in more details below (cf. 
Figs.~\ref{fig:Kitaev2Regimes}, \ref{fig:Z2fourRegimes}, \ref{Fig:U(1)Analytic}) there are combinations of the parameters $\{v_{Fi}, p_{Fi}\}$ for $i=1,2,3$, and of the soft frequency cut-off
that create regimes in which temperature scaling of the drag rate is qualitatively different.
In the next paragraphs we calculate the drag relaxation rate separately for a $\mathbb{Z}_2$ Kitaev QSL and both for a $\mathbb{Z}_2$ and a U(1) QSL together.
 
\subsection{$\mathbb{Z}_2$ QSL with Dirac excitations (Kitaev QSL)}
\label{sec:KitaevNoRelax}

\begin{figure}
\centering
\includegraphics[width=0.45\textwidth]{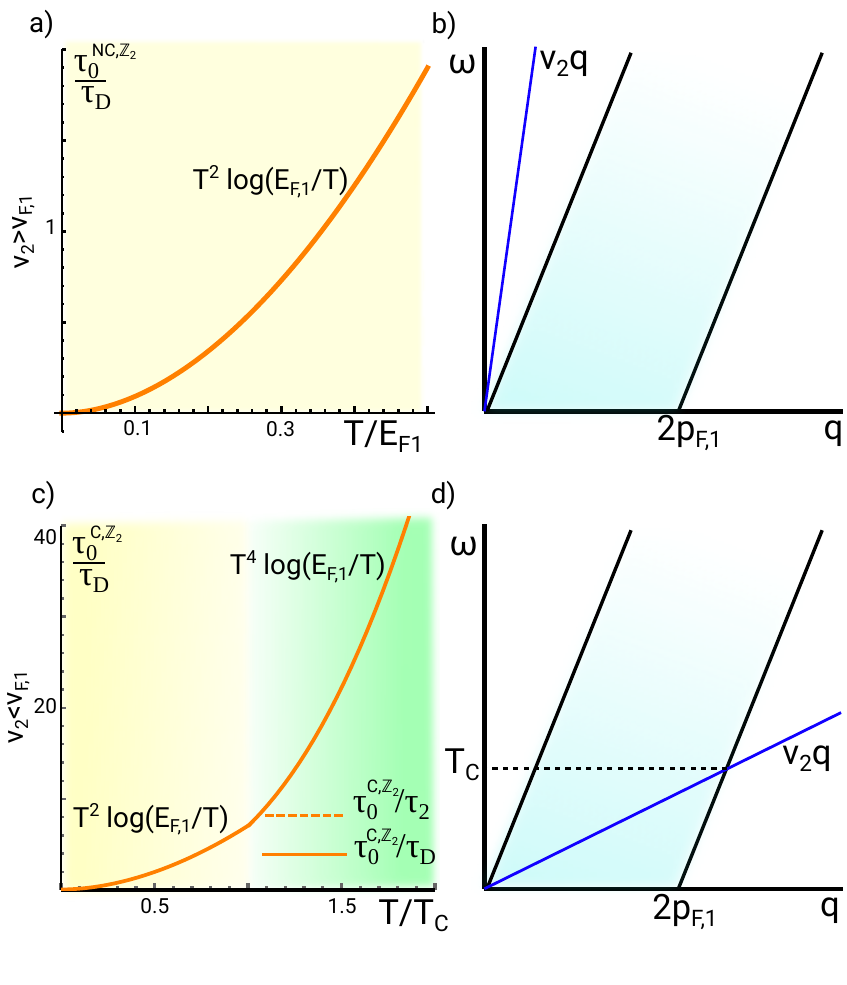}
\caption{Temperature dependence  
of the relaxation rate in the regime of fast spinons (panel a) and slow spinons (panel b) in a $\mathbb{Z}_2$ Kitaev QSL.  
In the case of fast spinons there is no crossover temperature below the Fermi temperature because in panel-b the line $\omega=2v_{F2}q$ never crosses the shaded area of particle-hole continuum. In contrast, for slow spinons, such a crossover takes place.} 
\label{fig:Kitaev2Regimes}
\end{figure}

In the case of  Kitaev QSLs we apply the formula from Eq. \eqref{eq:ALDrag} with the spin-correlator as in Eq.~\eqref{eq:DynStructKit}. As illustrated  in Fig.~\ref{fig:Kitaev2Regimes} there are  
two qualitatively different regimes, namely $v_{F2}>v_{F1}$ (\textit{fast spinons}) and $v_{F2}<v_{F1}$ (\textit{slow spinons}). The support of $\text{Im}[C^R(\v q,\omega)]$ at $\omega >v_{F2} q$, which is indicative of the continuum of particle-hole excitations about the Dirac point, overlaps with the particle-hole continuum $\omega < v_{F1} q$ of the Fermi-liquid like metallic bands only in the second case. This situation allows for a crossover temperature $T_C{\simeq 2 v_{F2} p_{F1}}$, above which the relative contribution of real processes dominates over virtual quantum processes. As shown in the Appendix \ref{app:kitaev}, in the limit of fast spinons we find 
\begin{subequations}
\begin{equation}
    \frac{1}{\tau_D}\simeq \frac{1}{\tau_0^{NC,K}}\left(\frac{T}{E_{F1}}\right)^2 \log{\left( \frac{E_{F1}}{T} \right)},
    \label{eq:tauDKitaevNC}
\end{equation}
with
\begin{equation}
    \frac{1}{\tau_0^{NC,K}}= \frac{4}{3}\left(\frac{J_K^2}{N}\right)^2 (C_K \nu_{{F1}} )^2T_C^2 E_{F1}.
    \label{eq:tau0KitaevNC}
\end{equation}
\end{subequations}
The result of Eq.~\eqref{eq:tauDKitaevNC} is shown in panel a) of Fig.~\ref{fig:Kitaev2Regimes}. In the limit of slow spinons we find instead
\begin{subequations}
\begin{equation}
    \frac{1}{\tau_D}\simeq \frac{1}{\tau_0^{C,K}}\left(\frac{T}{T_C} \right)^2\mathrm{max}\left[1,\left(\frac{T}{T_C}\right)^2\right]\log{\left(\frac{E_{F1}}{T}\right)}
    \label{eq:tauDKitaevC}
\end{equation}
and
\begin{equation}
  \frac{1}{\tau_0^{C,K}}=\left( \frac{T_C}{E_{F1}}\right)^2\frac{1}{\tau_0^{NC,K}}.
   \label{eq:tau0KitaevC}
\end{equation}
\end{subequations}
The result of Eq.~\eqref{eq:tauDKitaevC} is shown in panel c)  
of Fig.~\ref{fig:Kitaev2Regimes}.

 \subsection{$\mathbb{Z}_2$ and U(1) QSLs with Fermi surface}\label{sec:ALFSurface}

We now turn to the situation of QSLs with Fermi surface. In the case of
spinons coupled to a $\mathbb{Z}_2$ gapped gauge field, the dynamical spin-spin correlation function retains the Fermi liquid like behavior of the Lindhard function. It is a more nontrivial fact that even for U(1) spin-liquids the dynamical spin-spin correlation function is Fermi liquid-like for small $\omega$ and $q$. As such, using the Aslamazov-Larkin diagrammatic approach given by Eq.~\eqref{eq:ALDrag}, U(1) and $\mathbb Z_2$ QSL can mostly be treated at equal footing.

As in the Kitaev QSL there are two regimes, i.e. the  
limit of fast spinons ($v_{F2}>v_{F1}$) and the limit of slow spinons ($v_{F2}<v_{F1}$). In each of these two cases, one may distinguish the case of small ($p_{F2}<p_{F1}$) or large ($p_{F2}>p_{F1}$) spinon Fermi surface. The kinematic constraints for these cases are also illustrated in Fig.~\ref{fig:Z2fourRegimes} below.
 
For $\mathbb{Z}_2$ and U(1) QSLs a crossover temperature scale, $T_C \simeq 2 v_{\rm min} p_{\rm min}\ll E_{F1,2}$,
appears due to kinematic constraints in the cases of slow spinons with large Fermi surface, see Fig.~\ref{fig:Z2fourRegimes}. We found that in the limit of fast spinons
\begin{equation}
    \frac{1}{\tau_D}\simeq
\left(\frac{J_K^2}{N}\right)^2 \frac{(2\nu_{F2} \nu_{F1})^2}{E_{F1}}T^2 \log{\biggl(\frac{E_{F1}}{T}\biggr)},
    \label{eq:tauDZ2fast}
\end{equation}
while in the limit of slow spinons
\begin{equation}
    \begin{split}
    \frac{1}{\tau_D}\simeq &\frac{1}{4}
    \left(\frac{J_K^2}{N} \right)^2
    \frac{(\nu_{F1} \nu_{F2})^2}{E_{F1}} T^2\log{\left(\frac{E_{F1}}{T}\right)}\\
    \end{split}
    \label{eq:tauDZ2slow1}
\end{equation}
for $T<T_C$, and
\begin{equation}
    \frac{1}{\tau_D}\simeq \left( \frac{J_K^2}{N}\right)^2 \frac{(2\nu_{F2}\nu_{F1})^2}{ E_{F1}}T^2\log{\left( \frac{E_{F1}}{T} \right)}
    \label{eq:tauDZ2slow2}
\end{equation}
for $T>T_C$. Several comments are in order in relation to results presented in this section. First, the term $\log{\left(\frac{E_{F1}}{T}\right)}$ appears as a finite temperature regularization of the divergence at the edge of the integration domain. Generically it appears in the lifetime of 2D Fermi liquids and also for phonon-mediated drag~\cite{BonsagerMacDonald2000,Khveshchenko2000}. Second, the application of the above results to the case of U(1) QSLs is valid only for spinons with large Fermi surface. In the opposite case, the momentum integral extends beyond the limit of $q \ll p_{F2}$ for 
which the spin-correlation function is known to take a Fermi liquid form.  Third, for both U(1) and $\mathbb Z_2$ QSL we find that the $1/\tau_D \ll 1/\tau_2$ under realistic conditions. A posteriori it is thus evident that the physically relevant drag rate should be obtained in the limit of fast momentum relaxation, which is treated using Boltzmann transport theory in the next section.
 

\section{Theory with momentum relaxation in the QSL layer}
\label{sec:Boltzmann}

\subsection{General formalism}

{In this section, we present the effective Boltzmann treatment to account for intra-layer momentum relaxation in QSLs with Fermi surface. Within this formalism} there are three coupled equations that describe how momentum is relaxed in a two-step process from the passive layer to the active layer, see Fig.~\ref{fig:Schematics} b). 
We are interested in the stationary state in which the applied electric field in layer 1 generates a current that dissipates momentum within this 
layer due to temperature independent impurity scattering (assumed to be dominant) and transfers momentum to the QSL, that relaxes momentum within itself and transfers momentum to the second metallic sheet.
We assume the linear response regime to be valid (small fields applied, small deviations from equilibrium). 
This setup is described by the system of equations
\begin{subequations}\label{eq:system}
\begin{empheq}[left=\empheqlbrace]{align}
 -e\v {E}_1\cdot\frac{\partial f_1^{eq}}{\partial \v {p}_1} &  =-\frac{\delta f_1}{\tau_1} {+  I^{\rm{coll}}_{2{\rightarrow}1}}; \label{B1 1}
  \\
    0 & = -\frac{\delta f_2}{\tau_2} +  I^{\rm{coll}}_{1{\rightarrow} 2}+I^{\rm{coll}}_{3{\rightarrow}2}; \label{B1 b}
  \\
  0 & = 
-\frac{\delta f_3}{\tau_3} +  I^{\rm{coll}}_{2{\rightarrow}3}.\label{B1 c}
\end{empheq}
\label{eq:systemofequations}
\end{subequations}
Here, $f_{1,3} = f_{1,3}(\v  p)$ are the distribution functions in the metallic layers, and $f_{2} = f_{2}(\v  p)$ ($f_2 = f_2 (\hat p, \omega)$) is the distribution function (pseudo-distribution function) in the QSL in the case of a $\mathbb Z_2$ ($U(1)$) QSL with spinon Fermi surface. We treat intralayer momentum relaxation in the relaxation time approximation{, where $\delta f_i =f_i -  f_i^{\rm eq}$} with $i=1,2,3$. In our treatment the momentum relaxation rate within each layer including the QSL is an external phenomenological parameter that allows for the description in {the} 
variety of physical situations described in Eqs.~\eqref{eq:Z2Scattering rate}--\eqref{eq:U1ScatteringRate}. The additional collision integrals $ I^{\rm{coll}}_{j \rightarrow i}(\v {p}_i)$ account for the transfer of momentum between the layers and give rise to the drag resistivity. They are treated beyond relaxation time approximation and are derived in the next section. 
Linearizing the distribution function 
we can write the drag current as
\begin{equation}
    \v {j}_3=-\frac{e}{m_3}\int \frac{ d^2p_3}{(2\pi)^2}\delta f_3(\v  p_3)\v {p}_3.
    \label{eq:DefDragCurrent}
\end{equation}
Thus, the solution of Eqs.~\eqref{eq:systemofequations} and \eqref{eq:DefDragCurrent} lead to the calculation of $1/\tau_D$.

\subsection{Collision integral of the momentum transfer}
\label{sec:MomentumTransfer}

\begin{figure}
    \includegraphics[width=0.4\textwidth]{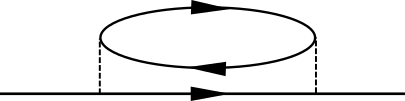}
    \caption{Self-energy diagram associated with the Kondo interaction between layers (dashed lines). Using the standard derivation of the Boltzmann equation from Keldysh technique, this diagram leads to Eq.~\eqref{eq:metal-qsl-linearized} describing the transfer of energy and momentum between layers.}
    \label{fig:sunsetdiagram}
\end{figure}

In this section, we summarize the calculation of the collision integral describing the momentum transfer between adjacent layers, leaving details to Appendix~\ref{app:KondoDerivation}. Using the Kondo coupling, Eq.~\eqref{KondoCouplingContinuum}, the Dyson equation in the Schwinger-Keldysh formalism, and a gradient expansion, the leading self-energy contribution due to Kondo coupling, Fig.~\ref{fig:sunsetdiagram}, leads to
\begin{align}
    I^{\rm{coll}}_{j \rightarrow i}=& \frac{J_K^2}{N}\int (d {p}_{i'}) (d{p}_j) (d{p}_{j'}){\delta({P}_i+{P}_{j}-{P}_{i'}-{P}_{j'})}\notag \\ 
    \times & \left\{ f^{\rm{eq}}_if^{\rm{eq}}_{j}[1-f^{\rm{eq}}_{i'}][1-f^{\rm{eq}}_{j'}](\psi_i+\psi_j-\psi_{i'}-\psi_{j'}) \right\}.
    \label{eq:metal-qsl-linearized}
\end{align}
In this expression we have absorbed a constant of order unity into $J_K$, and employed the linearization of the Boltzmann equation using
\begin{equation}
    \delta f_i=f_i^{\rm eq}(1-f_i^{\rm eq})\psi_i.
    \label{eq:deltafdefinition}
\end{equation}
The meaning of the $P_i$ entering the three dimensional delta function enforcing energy and  momentum conservation is explained below. 
For the non-linearized expression, see Eq.~\eqref{eq:metal-qsl} of the Appendix. 

It is important to stress several points concerning the notation and the fundamental distinction between a Fermi liquid and a nFL.
For the metal layer and the $\mathbb{Z}_2$ QSL
\begin{equation}
\int (d{p})(...)= \int \frac{d^2p}{(2\pi)^2}(...), \;  f^{\rm eq}_{{i}}=f^{eq}(\epsilon_{\v {p}_i}), \; {\psi_i = \psi(\v  p_i)}.
\label{def:misura1}    
\end{equation}
Consequently, the energy-momentum three-vector is $P_i = (\v p_i, \epsilon_{\v p _i})$.
In contrast, for the U(1) QSL layer, because of the compromised definition of quasi particles,
\begin{equation}
    \int (d{p})(...)= \nu \int d\epsilon \langle... \rangle_{{\hat p}},\; f^{eq}_i=f(\epsilon_i), \; {\psi_i = \psi(\hat p_i, \epsilon_i)},
    \label{def:misura2}
\end{equation}
where $\nu= m_2/2\pi$ is the density of states per spin and $\v  p_i = p_{F2} \hat p_i$, is implied everywhere, so that $P_i = (p_{F2} \hat p_i, \epsilon_i)$, in this case. 

\subsection{General solution for the drag rate}

The system of Eqs.~\eqref{eq:systemofequations}--\eqref{eq:DefDragCurrent} can be manipulated to derive a general formula for the drag conductivity (see Appendix \ref{app:GeneralFormula} for details):
\begin{align}
    \frac{1}{\tau_D}&=-\frac{(2\pi)^5}{n_1 n_3} \frac{\tau_2}{2T}\left (\frac{J_K^2}{N}\right)^2 \int dQdQ' (\v {q}\cdot \hat{E}_1) (\v {q}'\cdot \hat{E}_1) \notag \\
    &\times  \frac{\mathrm{Im}[\Pi_3^{\mathrm{R}}(Q)] \mathrm{Im}[\Pi_1^{\mathrm{R}}(Q')] I_2(Q,Q')}{\sinh{\bigl( \frac{\omega}{2T} \bigr)} \sinh{\bigl( \frac{\omega'}{2T} \bigr)} \sinh{\bigl( \frac{\omega+\omega'}{2T} \bigr)}},
\label{eq:DragGeneralMainText}
\end{align}
with $Q=(\v q,\omega)$ and $dQ=\frac{d^2q}{(2\pi)^2}\frac{d\omega}{2\pi}$,
and we have defined
\begin{equation}
\begin{split}
I_2(Q,Q')&=\int (d{p}_{2})(d{p}_{2'})(d{p}_{2''})  \delta^{(3)}({P_{2''}-P_{2'}+Q})\times\\
    &\times \delta^{(3)}({P_{2''}-P_{2}-Q'}) (f_2-f_{2'}).
\end{split}
\label{eq:I2DefMainText}
\end{equation}
Apart from the behavior of $\tau_2$, the integral
$I_2(Q,Q')$ encodes the main difference between the the $\mathbb{Z}_2$ and the U(1) QSL. 

\subsection{$\mathbb Z_2$ QSL with Fermi surface}{\label{sec:Z2Boltzmann}}

\begin{figure*}
\centering
\includegraphics[width=\textwidth]{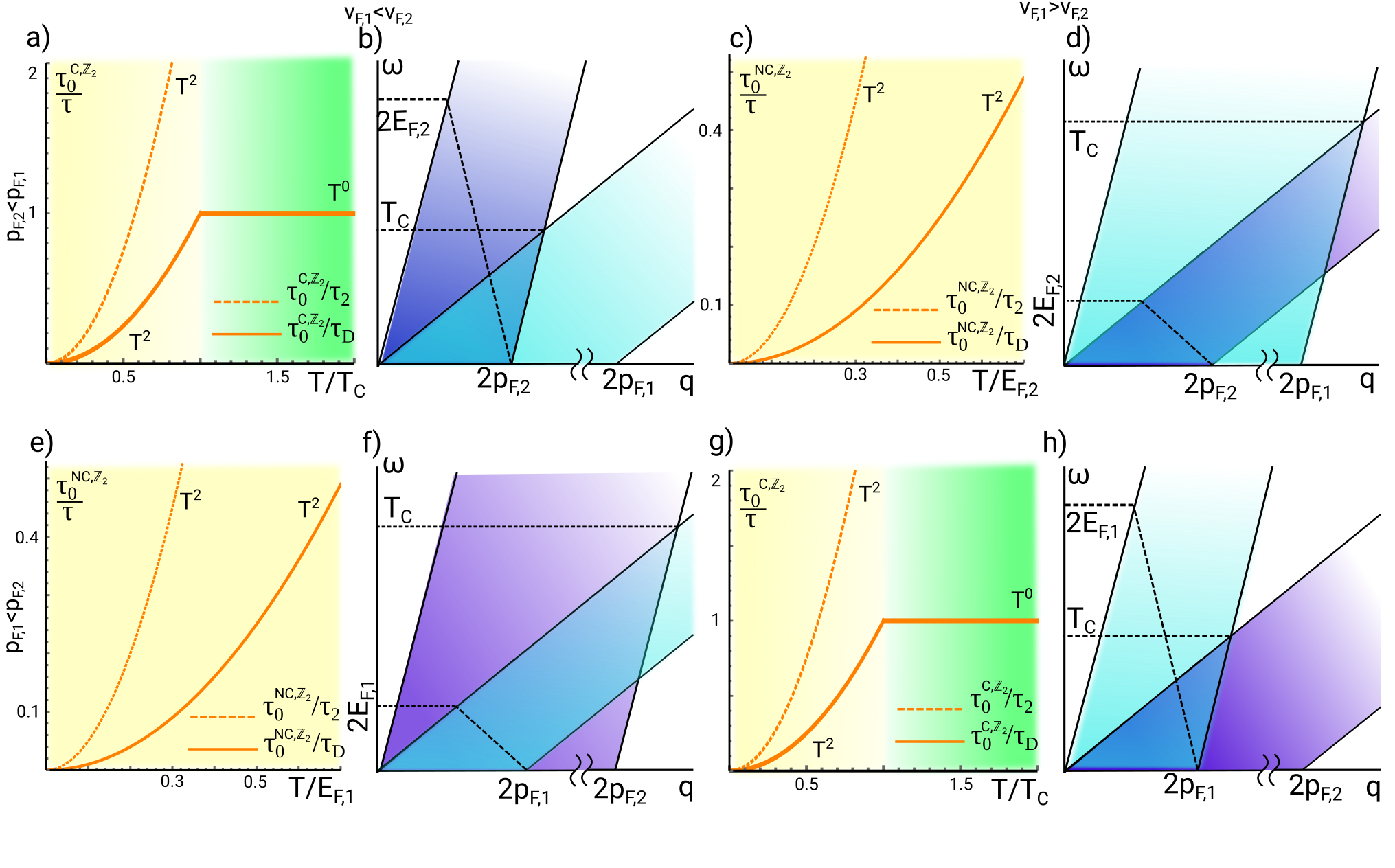}
\caption{Temperature dependence of the spinon-mediated drag in $\mathbb Z_2$ QSLs, Eqs.~\eqref{eq:tauDZ2BoltzNC}--\eqref{eq:tauDZ2BoltzC}, and regions of kinematic constraints for the momentum transfer.
The combination of Fermi velocities and momenta of spinons and electrons leads to four regimes (particle-hole continua in the metallic lead (QSL) are shaded light blue (dark purple)).
The panels b) and h) are symmetric under the exchange $1\leftrightarrow2$ and differ from each other only for the value of $\tau_0^{{C, \mathbb Z_2}}$, whose definition is generally given by Eq.~\eqref{eq:tau0Z2C} and $T_C=2v_{\mathrm{min}}p_{\mathrm{min}}$. Finally, panels b) and c) are symmetric (for $1 \leftrightarrow 2$) and only differ for the value of $\tau_0^{{NC, \mathbb Z_2}}$ defined in Eq.~\eqref{eq:tau0Z2NC}.}
\label{fig:Z2fourRegimes}
\end{figure*}

In the case of the $\mathbb{Z}_2$ QSL with the Fermi surface we calculate the object $I_2(Q,Q')$ to be 
\begin{equation}
    \begin{split}
        &{\langle I_2(Q,Q')({\v{q}\cdot \hat{E}_1})({\v{q}'\cdot \hat{E}_1})
    \rangle_{\hat{q},\hat{q}'}}={-}\frac{4T \sinh{\bigl(\frac{\omega+\omega'}{2T} \bigr)}}{(2\pi)^6\omega\omega'E_{F2}}
    \\
        &\times \biggl( \omega+\frac{q^2}{2m_2} \biggr)  \biggl( \omega'+\frac{q'^2}{2m_2} \biggr) {\rm{Im}}\left[\Pi_2^R(\v {q},\omega)\right]{\rm{Im}}\left[\Pi_2^R(\v {q}',\omega')\right],\\
    \end{split}
    \label{eq:I2 for Z2}
\end{equation}
which holds in the regime
\begin{equation}
\frac{\omega}{v_{\rm{F}}}\left(1+\frac{T}{2E_F}\right)<q<\left (2p_{\rm{F}}-\frac{\omega}{v_{\rm{F}}}\right)\left(1-\frac{T}{2E_F}\right)
\label{eq:condition_int_boundary2}
\end{equation}
that defines the dominant contribution to the integrals entering the drag rate.

The insertion of Eq.~\eqref{eq:I2 for Z2} into the general formula for the drag rate, Eq.~\eqref{eq:DragGeneralMainText}, implies that the domain of integration over $Q, Q'$ is bounded by the overlap of particle-hole continua in both metallic and QSL Fermi liquids, see Fig.~\ref{fig:Z2fourRegimes}. Physically, this means that only real scattering processes between particle-hole pairs of adjacent layers allow for momentum transfer. As a consequence, the combinations of the parameters $p_{Fi}$ and $v_{Fi}$ give rise to four 
regimes that we indicate formally by the Cartesian product $(v_{F1}>v_{F2}, v_{F1}<v_{F2})\times(p_{F1}>p_{F2}, p_{F1}<p_{F2})$. This is analogous to the regimes introduced in Sec.~\ref{sec:ALFSurface}, but in contrast to the present discussion, we considered there the limit of absent momentum relaxation in the QSL and thereby also included virtual particle-hole excitations in the QSL.

We first discuss the limit {$({v_{F1}}>{ v_{F2}}  \land {p_{F1}}>{ p_{F2}}) \lor ({v_{F1}}<{ v_{F2}}  \land {p_{F1}}<{ p_{F2}})$}, corresponding to panels d) and f) of Fig.~\ref{fig:Z2fourRegimes}. 
As we assume the low-temperature degenerate Fermi gas limit
of either the metallic plates or the QSL, the minimal Fermi energy  $E_{\text{min}}=\text{min}\left(E_{F1},E_{F2} \right)$ bounds the temperature limiting the energy integration. Hence, the drag rate in the symmetric case is
\begin{subequations}
\begin{equation}
    \frac{1}{\tau_D}= \frac{1}{\tau_0^{NC,\mathbb{Z}_2}}\left(\frac{T}{E_{\text{min}}}\right)^2,
    \label{eq:tauDZ2BoltzNC}
\end{equation}
with
\begin{equation}
    \frac{1}{\tau_0^{NC,\mathbb{Z}_2}}=\frac{1}{4\pi^7}\left(\frac{J_K^2}{N}\right)^2 \left(\frac{E^2_{\text{min}}}{E_{F1}E_{F2}}\right)\frac{\nu_{2} p_{\text{min}}^4}{v_{F1}^2},
    \label{eq:tau0Z2NC}
\end{equation}
\end{subequations}
as shown in panels c) and e) of Fig.~\ref{fig:Z2fourRegimes}. 

We discuss next the regimes  
$(v_{F1}>v_{F2}  \land p_{F2}<p_{F1}) \lor (v_{F1}<v_{F2}  \land p_{F1}> p_{F2})$. As shown in panels b) and h) Fig.~\ref{fig:Z2fourRegimes} 
the crossover temperature $T_C \simeq 2v_{\text{min}}p_{\text{min}}$ [with $v_{\text{min}}=\min\left( v_{F1}, v_{F2}\right)$ and $p_{\text{min}}=\min\left( p_{F1}, p_{F2}\right) $] emerges. For $T<T_C$, the frequency integration is bounded by $T$, while the overlap of particle-hole continua limits the $\omega$ integral at $T_C$ for $T_C< T$. Importantly, the crossover temperature scale is below $E_{\rm min}$ and thus relevant for the regime of applicability of our theory. The drag rate in these regimes is thus   
{
\begin{subequations}
\begin{equation}
    \frac{1}{\tau_D}=\frac{1}{\tau_0^{C,\mathbb{Z}_2}} \left(\frac{T}{T_C} \right)^2\mathrm{min}\left[1,\left( \frac{T_C}{T}\right)^2\right],
    \label{eq:tauDZ2BoltzC}
\end{equation}
with
\begin{equation}
    \frac{1}{\tau^{C,\mathbb{Z}_2}_0}=\pi \left( \frac{E_{\text{min}}}{T_C}\right)^2 \frac{1}{\tau^{NC,\mathbb{Z}_2}_0}.
    \label{eq:tau0Z2C}
\end{equation}
\end{subequations}}
These results are plotted in panels a) and g) in Fig.~\ref{fig:Z2fourRegimes}.

\subsection{{U(1) QSL with Fermi surface}}{\label{sec:U(1)Boltzmann}}

\begin{figure}
\centering
\includegraphics[width=0.5\textwidth]{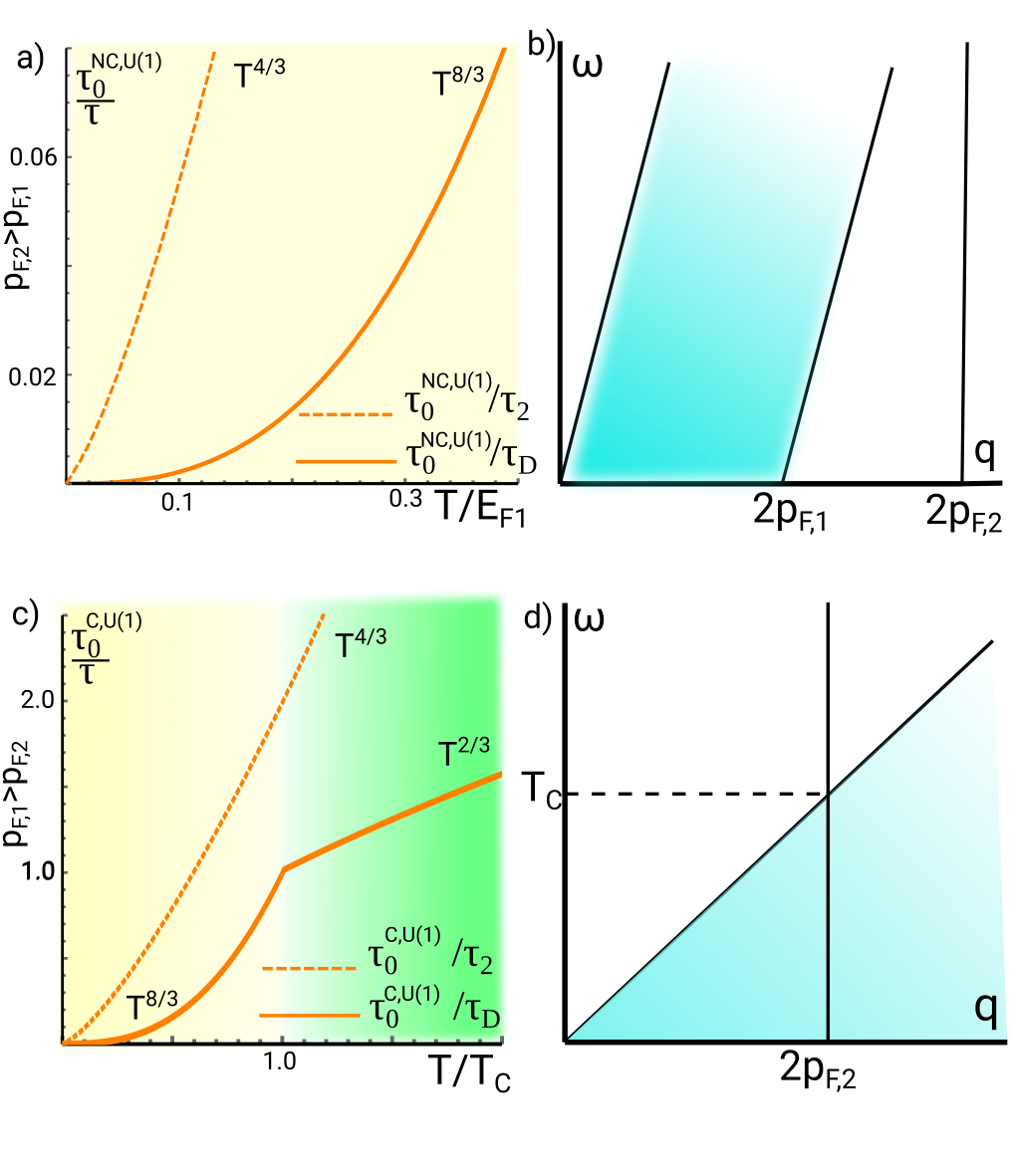}
\caption{Temperature dependence of the spinon-mediated drag in $U(1)$ QSLs, per Eqs.~\eqref{eq:tauDU(1)NC}--\eqref{eq:tauDU(1)C}, and kinematic constraints for the momentum transfer. 
Panels a) and c) illustrate that  
$\tau_2^{-1}\gg \tau_D^{-1}$, meaning that momentum relaxation in the QSL is faster than interlayer momentum transfer. The latter underpins applicability of the semiclassical Boltzmann approach.}
\label{Fig:U(1)Analytic}
\end{figure}

In the case of the U(1) QSL we find
\begin{equation}
\begin{split}
    & {\bigl \langle I_2(Q,Q')(\v {q}\cdot\hat{E_1})(\v {q}'\cdot\hat {E}_1)\bigr\rangle_{\hat{q},\hat{q}}}=\\
    &=-\frac{\nu^3_2}{\pi p_{F2}^2}{({\omega+\omega'})}\frac{\theta(2p_{F2}-q)q}{\sqrt{(2p_{F2})^2-q^2}} \frac{\theta(2p_{F2}-q') q'}{\sqrt{(2p_{F2})^2-q'^2}},
    \end{split}
    \label{eq:IIntegralU1}
\end{equation}

Insertion of Eq.~\eqref{eq:IIntegralU1} into Eq.~\eqref{eq:DragGeneralMainText} leads to the general equation for the drag rate in the case of U(1) QSL. As opposed to the $\mathbb Z_2$ case, Eq.~\eqref{eq:I2 for Z2}, Eq.~\eqref{eq:IIntegralU1} only contains a
sharp momentum cutoff $\theta(2p_{F2}-q)$ that is a consequence of the fact that for this pseudo quasiparticles the momentum is pinned on the Fermi surface but no bound in the frequency integrals over $\omega$, $\omega'$. This is a consequence of the diffuse nature of the fermionic spectral weight in a nFL with self-energy $\Sigma \sim \epsilon^{2/3}$, which allows for real scattering processes even when energy-conservation would be violated in a Fermi liquid with sharp quasi-particles.

The altered kinematic constraints, see Fig.~\ref{Fig:U(1)Analytic}, for the nFL lead to qualitatively different regimes and behaviors of $\tau_D^{-1}(T)$, because in this case the velocity $v_{F2}$ does not play any role. We can therefore distinguish two limits: (i) $p_{F2}>p_{F1}$ in which case, as shown in Fig.~\ref{Fig:U(1)Analytic} {b)}, the $\omega$-integration is bounded by $T<E_{F1}$, and there is no other relevant energy scale; (ii) the limit $p_{F2}< p_{F1}$, in which case the integration can be bounded by the domain integration.

In the symmetric regime for $p_{F2}\gg p_{F1}$ the drag rate is found in the form 
\begin{equation}
    \frac{1}{\tau_D}=\frac{1}{\tau_0^{NC,U(1)}}\left(\frac{T}{E_{F1}}\right)^{\frac{8}{3}},
    \label{eq:tauDU(1)NC}
\end{equation}
with
\begin{equation}
    \frac{1}{\tau_0^{NC,U(1)}}=\frac{128}{\pi^2}\left(\frac{J_K^2}{N} \right)^2\frac{\nu_1 \nu_3 \nu_2^3}{p_{F2}^2}\frac{E_{F1}^4}{T_{C}^2} \left( \frac{E_{F2}}{E_{F1}}\right)^{1/3}.
    \label{eq:tau0U(1)NC}
\end{equation}
This result is plotted in panel a) of Fig.~\ref{Fig:U(1)Analytic}.
In the limit of $p_{F2}\ll p_{F1}$ the temperature dependence depends on
a crossover temperature $T_C=2v_{F1}p_{F,}$ and the relaxation rate is given by
\begin{equation}
 \frac{1}{\tau_D}=\frac{1}{\tau_0^{C,U(1)}}\left(\frac{T}{T_C}\right)^{\frac{8}{3}}\mathrm{min}\left(1,\frac{T_{C}}{T}\right)^2,
   \label{eq:tauDU(1)C}
\end{equation}
where the prefactor is given by 
 \begin{equation}
    \frac{1}{\tau_{0}^{C,U(1)}}=\frac{1}{\tau_{0}^{NC,U(1)}}\left(\frac{1}{4}\right)\left(
    \frac{p_{F1}}{p_{F2}}\right)^2\left(\frac{T_C}{E_{F1} }\right)^{\frac{14}{3}}.
    \label{eq:tau0U(1)NC-C}
\end{equation}


\section{Summary, Discussion, and outlook}

In summary, we have derived the spinon-mediated induced drag rate for heterostructures as displayed in Fig.~\ref{fig:Schematics} a) for a three types of QSLs: Kitaev QSL beyond the integrable limit; $\mathbb Z_2$ QSL with Fermi surface; U(1) QSL with Fermi surface. We have distinguished the situation of the slow momentum relaxation within the QSL layer (as compared to the drag rate) from the situation of fast momentum relaxation. In the first case, we have derived the drag rate using Aslamazov-Larkin diagrammatics, while we employed a Boltzmann-like description for the second case. We conclude that the latter limit appears experimentally more realistic (see dashed curves in Figs.~\ref{fig:Z2fourRegimes},\ref{Fig:U(1)Analytic}). The semiclassical Boltzmann approach assumes the Fermi wavelength to be short as compared to the mean free path, which is directly applicable only to QSLs with a Fermi surface in a proper range of parameters. In contrast to earlier studies \cite{ChenXie2013,ChattarjeeSachdev2015,Okamoto2016,WermanBerg2018,AftergoodTakei2020,ZhuangMarston2021}, the transport phenomena discussed here are purely electrical, and do not rely on the spin injection through spin-orbit interaction in the leads.

\begin{figure}
    \centering
    \includegraphics[width = 0.45\textwidth]{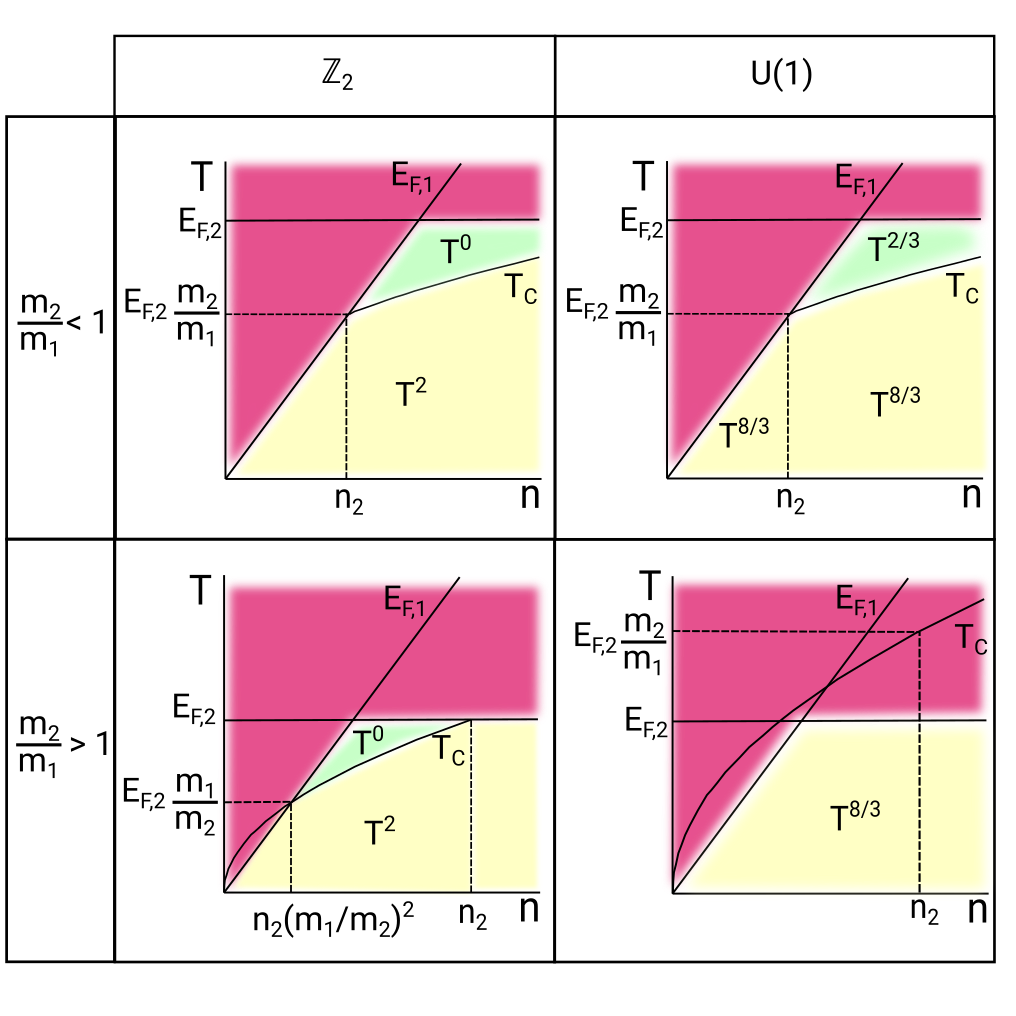}
    \caption{{The figure shows the scaling behavior of $\tau_D^{-1}$ as a function of temperature and charge carriers in the metallic layers and in the symmetric regime, in the case of the metallic plates having a parabolic dispersion.
    Only  the results for the $\mathbb{Z}_2$ and U(1) QSL with Fermi surface are reported because the notion of $m_2$ does not exist in the context of a Kitaev QSL}.
    }
    \label{fig:tauDSummary}
\end{figure}

We now summarize and discuss main results as displayed in Figs.~\ref{fig:Schematics} c) (for graphene) and \ref{fig:tauDSummary} (for 2DEGs with parabolic dispersion) as a function of experimentally tunable parameters, namely temperature and charge carriers density $n\sim p_{F1}^2$ in the metallic leads.  
The difference between the two figures is due to the independence (square-root dependence) of the Fermi velocity on the carrier density in graphene (in a 2DEG with parabolic dispersion). Note that in all our Boltzmann calculations we assumed a quadratic dispersion both for the metallic layers and for the QSL layer. However, as previously discussed, 
the results most crucially depend on the Fermi-liquid like particle-hole continua of the metallic leads. At small momenta excitations, these are the same for graphene or 2DEGs with parabolic dispersion. Therefore, our calculations are expected to qualitatively describe metallic layers made of doped graphene, as well (see also Sec.~\ref{sec:DragBasics}).

Except for the Kitaev QSL, our approach assumes a degenerate Fermi gas 
$T\ll E_F$ both for the metallic leads and the spinons. The purple region in all the panels in Figs.~\ref{fig:Schematics} c) and \ref{fig:tauDSummary} is beyond this assumption.  
The remaining areas of the parameter space contain characteristic crossover temperature scales, above which the power exponent of the temperature dependence of the drag rate changes. This is illustrated by means of the color coding in Figs.~\ref{fig:Schematics} c) and \ref{fig:tauDSummary} and a consequence of the kinematic constraints of energy and momentum transfer.

Specifically, in the case of the Kitaev QSL, a crossover in temperature occurs $T_C=2v_{F2}p_{F1}$ below the red region in the case of \textit{slow spinons} ($v_{F2}<v_{F1}$). In a $\mathbb{Z}_2$ QSL with Fermi surface there is a crossover temperature $T_C=2v_{\text{min}}p_{\text{min}}$ in both the cases  
of fast and slow spinons. Note that for graphene leads, Fig.~\ref{fig:Schematics} c), in the case of fast spinons $T_C=2v_{F1}p_{F2}$, so that the characteristic crossover temperature is not a function of density $n$. 
Finally, for the U(1) QSL with Fermi surface the drag relaxation time has a crossover temperature $T_C=2v_{F1}p_{F2}$ only in the limit of fast spinons. We stress that this is a characteristic difference as compared to the case of the $\mathbb{Z}_2$ QSL and ultimately a consequence of the 
smeared fermionic spectral weight due to the non-Fermi-liquid behavior of the U(1) QSL. Another key distinction of the U(1) QSL are the fractional powers in the temperature dependence, which reflect the momentum relaxation rate $1/\tau_{2} \sim T^{4/3}$ inside the QSL.

We conclude with a discussion of experimental consequences of this study. First, as in all 2DEG double-layers, Coulomb interaction provides the leading mechanism for drag current generation. A way to experimentally suppress the contribution of Coulomb interaction without affecting the spinon drag contribution is to embed the heterostructure in the environment of a strong dielectric, such as SrTiO$_3$, which are often employed experimentally for gating purposes. Second, the spinon mediated drag is a fourth order effect in Kondo coupling $J_K$ and thus naively small in amplitude. However, the reference energy scale determining the drag rate is given by a combination of $E_{F1}$ and the microscopic energy scale of the QSL $J$. While the former can be reduced by gating whereby the spinon induced drag increases, the ratio $J_K/J$ determines whether the heterostructure is in a topologically ordered, fractionalized Fermi liquid~\cite{SenthilVojtaSachdev2003} (FL$^*$) with small Fermi surface (small $J_K/J$) or a Kondo-screened phase of a topologically trivial Fermi liquid (FL) with large Fermi-surface (large $J_K/J$). Importantly, also the second regime appears experimentally realistic~\cite{MashhadiKern2019}.

We conclude with a speculation about the behavior of the drag rate as a function of $J_K/J$ as the FL$^*$ to FL transition is approached and hypothesize that the $1/\tau_D \sim J_K^4$ increase levels off at the transition and gives way to a regime of perfect drag, akin to the situation of perfect drag in the presence of an exciton condensate~\cite{NandiWest2012}. Indeed, the mean field description of the Kondo-screened phase implies that only the symmetrized wavefunction of electrons of layers 1 and 3 form the FL with large Fermi surface, while the antisymmetric wave function of electrons constitutes an independent band with small Fermi surface. The homogeneous dc field $\v E_1$ does not allow for interband coupling and has support in both bands, the same is true for the current operator $\v j_3$. By consequence, electrons in both bands are accelerated and equally contribute to drag and transport current, ultimately leading to the perfect drag phenomenon (same local and non-local conductances). It will be an interesting problem for the future to study this effect in more detail, as well as the impact of magnetic field on the spinon mediated drag.

\acknowledgments

We gratefully acknowledge useful discussions with Inti Sodemann, Ciaran Hickey, Marko Burghard. A.L. is grateful to the Max Planck Institute for Solid State Research for hospitality, where this work was initiated, and to the Alexander von Humboldt Foundation for the financial support.   

\appendix

\section{Limit of absent momentum relaxation within QSL layer: {Aslamazov-Larkin-Diagram}}

In this appendix, we present details complementing discussion in  Sec.~\ref{sec:NoMomentumRelax}. As argued in the main text, 
when the condition $\tau_{2}^{-1}\ll \tau_{\rm{D}}^{-1}$ is satisfied,
$\tau_{D}^{-1}$ is analogous to the Coulomb drag one \cite{NarozhnhyLevchenko2016}, as determined by the Aslamazov-Larkin diagram (see Eq.~\eqref{eq:ALDrag}).

In the case of QSL with Fermi surface and for small $\omega$ and $q$, the spin correlation function is proportional to 
the free electron polarization operator 
\begin{equation}
    \Pi_{{i}}^{\rm{R}}(\v {q},\omega)=\int (d{k})\,
    \frac{f^{eq}(\epsilon_{\v {k}})-f^{eq}(\epsilon_{\v {k}+\v {q}})}{\epsilon_{\v {k}}-\epsilon_{\v {k}+\v {q}}-\omega+i0^+},
    \label{A2}
\end{equation}
with $\epsilon_{\v {k}}=|\v {k}|^2/2m - E_{Fi}$, $E_{Fi}$ being the Fermi energy, with $f^{eq}$ being the Fermi-Dirac equilibrium distribution function.

\subsection{Properties of the polarization operator}\label{AppendixPolOp}

Despite the strong coupling nature and the nFL behavior of the U(1) QSL, it is still possible to show that for small $\omega$ and $q$ the polarization operator for the $\mathbb{Z}_2$ and the U(1) QSLs are essentially the same as the free electron polarization operator~\cite{AltshulerMillis1994}. Furthermore, if $\Pi^R(\v {q},\omega)_{T= 0,E_F=E}$ is a smooth function in an interval with a width of $\delta E\simeq{T}$ around $E=E_F$, we can use the zero temperature expression of the polarization operator also in the finite temperature situation. At zero temperature ${T}=0$ the polarization operator is~\cite{GiulianiVignale2005}
\begin{equation}
\begin{split}
&\text{Im}\left[\Pi^R(\v {q},\omega)\right] = -\nu\frac{m}{q^2}\{\theta((v_F q)^2- \omega_{-}^2)\sqrt{(v_F q)^2- \omega_{-}^2} \\
&-\theta((v_F q)^2- \omega_{+}^2)\sqrt{(v_F q)^2- \omega_{+}^2}\}; \\
&\text{Re}\left[ \Pi^R(\v {q},\omega)\right]=-\nu\biggl\{1+\nu\frac{m}{q^2}\{
\text{sign}(\omega_{-})\theta(\omega_{-}^2-(v_F q)^2)\times\\
&\times\sqrt{\omega_{-}^2-(v_F q)^2}
-\text{sign}(\omega_{+})\theta(\omega_{+}^2-(v_F q)^2)\sqrt{\omega_{+}^2-(v_F q)^2}
\} \biggr\} ,
\end{split}
\label{T=0 PolOp}
\end{equation}
where $\omega_{\pm}=\left(\omega \pm \frac{q^2}{2m} \right)$ and $\nu$ is the density of states at the Fermi surface. To define quantitatively the applicability condition of $T\to0$ expressions one turns to the imaginary part of the finite temperature polarization operator that can be written as follows 
\begin{align}
{\rm{Im}}[\Pi^{R}(\mathbf{q},\omega)]&= -\nu\frac{m}{q^2}\int_{-\infty}^{+\infty} dE  \left(\frac{\partial f^{eq}}{\partial E}\right) \times \notag \\
& \times \biggl\{ \sqrt{(v_{F}q)^2 -
\omega_{-}^2 + \frac{2q^2}{m}(E-E_{F})} \notag\\
&-\sqrt{(v_{\rm{F}}q)^2-\omega_{+}^2+ \frac{2q^2}{m}(E-E_{F})} \biggr\}.
\end{align}
Here, it is assumed that the argument of the square roots is positive (otherwise the expression vanishes per definition).

\begin{figure}[]
\centering
\includegraphics[width=0.5\textwidth]{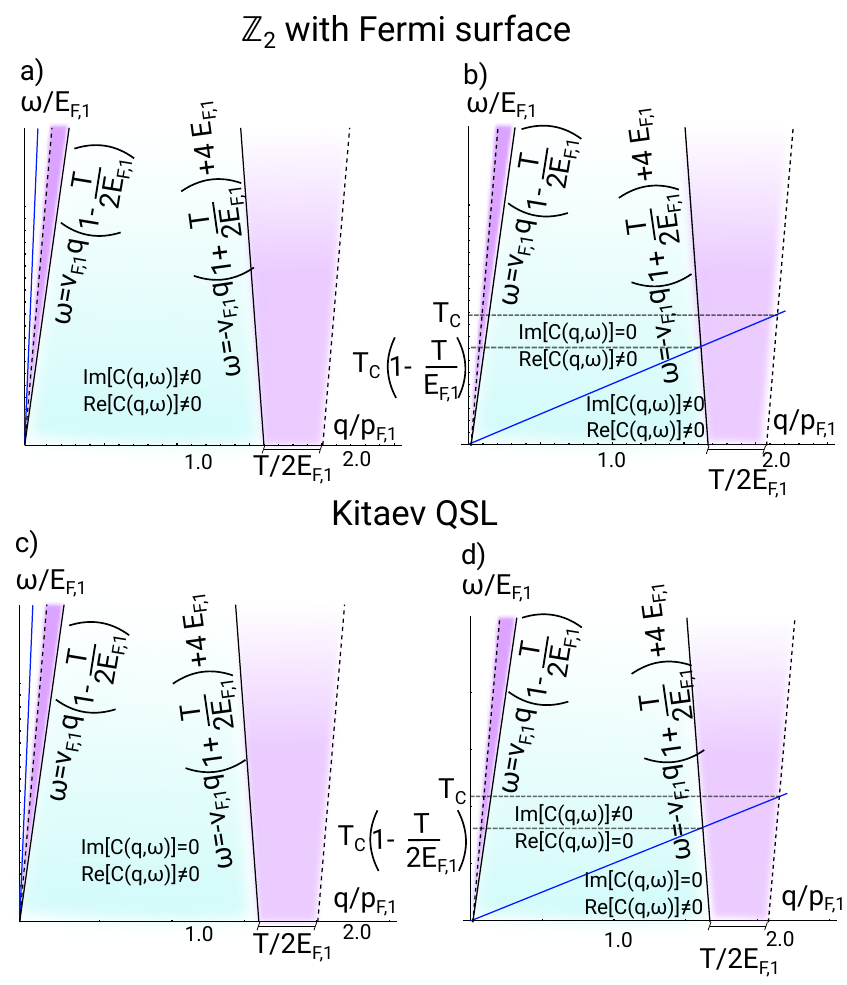}
\caption{Integration domain used for integrals in Eq.~\eqref{eq:ALDrag} (light blue shading). Solid lines are given by Eq.~\eqref{eq:condition_int_boundary2}. The areas in purple are discarded, their  contributions are higher orders of $\frac{T}{E_F}$. Panels a) and b): $\mathbb{Z}_2$ QSL with Fermi surface in the case of fast and slow spinons respectively. As shown in panel b), in the case of slow spinons there is a crossover in the behavior of $|C^R(\v q,\omega)|$.
The same considerations old true for the Kitaev QSL (panels c) and d)) with a difference for the crossover of the spin correlation function over the line $\omega=v_{F2} q$.  }
\label{fig:Z2Detailed}
\end{figure}

We now comment on the role of temperature in determining the boundaries of the integral support. We note that the term $(v_{F}q)^2 -
\omega_{\pm}^2 $ would vanish in proximity of the boundary of the support of the imaginary part of the polarization operator. The only term that keeps the square root smooth is $ \frac{2q^2}{m}(E-E_F)$ which is bounded by $T$, so that Eq.~\eqref{T=0 PolOp} will always be valid provided that the condition
\begin{equation}
    (v_{F}q)^2-\omega_{\pm}^2>\frac{2q^2}{m}T
    \label{A7}
\end{equation}
is satisfied. This leads to a restriction in the integration over $q$ given by Eq.~\eqref{eq:condition_int_boundary2}. This integration domain is represented in blue in Fig.~\ref{fig:Z2Detailed}. 
It should be pointed out that in this regime the expansion of $\Pi_{T=0}(\mathbf{q},\omega)$ coincides with $\Pi_{T\ne0}(\mathbf{q},\omega)$. Also $\Pi_{T=0}^R(\mathbf{q},\omega)$ can be expanded if $\omega_{\pm}^2<<(v_F q)^2\longrightarrow |\omega_{\pm}|\ll v_F q$. 
This condition is less restrictive than \eqref{A7}, so since we limit the integration in this domain, we can expand ${\mathrm{Im}}[\Pi^R(\mathbf{q},\omega)]$ to obtain
\begin{equation}
    \mathrm{Im}[\Pi^R(\mathbf{q},\omega)]\simeq \frac{\nu\omega}{\sqrt{(v_F q)^2-\bigl( \omega^2+\bigl(\frac{q^2}{2m}\bigr)^2 \bigr)}}.
    \label{A8}
\end{equation}
where we have dropped to subscript $T=0$ because from now on we will deal only with the $T=0$ polarization operator. The expression is valid for the whole integration regime and especially for the QSL we will make large use of the approximation $q\ll p_{F}$ that leads to 
\begin{equation}
    \Pi^R(\mathbf{q},\omega)\simeq 
    \nu \left( 1 + i\frac{\omega}{\sqrt{(v_F q)^2-\omega^2}} \right).
    \label{A9}
\end{equation}

\subsection{$\mathbb Z_2$ QSL with Dirac excitations (Kitaev QSL)}\label{app:kitaev}

We rewrite Eq.~\eqref{eq:ALDrag} as follows 
\begin{equation}
    \frac{1}{\tau_D}={\left(\frac{J_K^2}{N}\right)^2}\frac{1}{ p_{F1}^2 m T} \int_0^{\infty}\frac{d\omega}{\sinh^2{\bigl( \frac{\omega}{2T} \bigr)}}I_q(\omega)
    \label{Kitaev21}
\end{equation}
and consider separately limits of slow and fast spinons as discussed in the main text.

\subsubsection{Limit of fast spinons}

We start with the case $v_{F2}>v_{F1}$ in which {$I_q(\omega) = I^{{NC}}_q(\omega)$,}
\begin{equation}
    \begin{split}
    &I^{{NC}}_q(\omega)=\int dq q^3 \left( \frac{(\nu_{1}\omega)^2}{(v_{F1}q)^2-(\frac{q^2}{2m})^2-\omega^2} \right)\times \\ 
    &\times C_K^2((v_{F2}q)^2-\omega^2)=(C_K \nu_{1})^2(\omega^2 v_{F2}^2 I_1-\omega^4 I_2),
    \label{Kitaev22}
    \end{split}
\end{equation}
here the NC stands for non-crossover, indicating that the crossover line is outside of the integration domain, and
\begin{equation}
 I_1=\int_{\frac{\omega}{v_{F1}}\left(1+\frac{T}{2E_{F1}}\right)}^{-\frac{\omega}{v_{F1}}\left(1-\frac{T}{2E_{F1}}\right)+2p_{F1}\left(1-\frac{T}{2E_{F1}}\right)} 
\frac{dq q^5}{(v_{F1}q)^2-(\frac{q^2}{2m})^2-\omega^2}, 
\end{equation}

\begin{equation} I_2=\int_{\frac{\omega}{v_{F1}}\left(1+\frac{T}{2E_{F1}}\right)}^{-\frac{\omega}{v_{F1}}\left(1-\frac{T}{2E_{F1}}\right)+2p_{F1}\left(1-\frac{T}{2E_{F1}}\right)}  
\frac{dq    q^3}{(v_{F1}q)^2-(\frac{q^2}{2m})^2-\omega^2}.
\end{equation}
The integral $I_1$ can be evaluated to be 
\begin{equation}
    I_1=8 p_{F1}^2 m^2 \left[\log{\left( \frac{E_{F1}}{T}\right)-1-\log{\left( 1+\frac{\omega}{2T}\right)}}\right],
    \label{Kitaev23}
\end{equation}
while the integral $I_2$ can be calculated to be
\begin{equation}
    I_2=2  m^2 \left[\log{\left( \frac{E_{F1}}{T}\right)-\log{\left( 1+\frac{\omega}{2T}\right)}}\right].
    \label{Kitaev24}
\end{equation}
Integration over $\omega$ of Eq.~\eqref{Kitaev22} yields the results presented in Eq.~\eqref{eq:tauDKitaevNC} and \eqref{eq:tau0KitaevNC} and shown in panel a) of Fig.~\ref{fig:Kitaev2Regimes}.

\subsubsection{Limit of slow spinons}

We turn now to the regime in which  $v_{F2}<v_{F1}$ which contains a typical crossover temperature $T_C=2v_{F2}p_{F2}$ and frequency
\begin{equation}
    \omega_C= T_C\left(1-\frac{T}{2E_{F1}} \right),
    \label{eq:def omegaC}
\end{equation}
see Fig.~\ref{fig:Z2Detailed} d). The $\omega$ integral appears exactly in the same form as \eqref{Kitaev21}, but this time we write
\begin{equation}
   \begin{split}
    &I^{{C}}_q=
    C_K^2 [\omega^4 I_3+ \omega^2 I_4],
    \end{split}
    \label{Kitaev26}
\end{equation}
with
\begin{align}
    I_3&=\int_{\frac{\omega}{v_{F1}}\left(1 + \frac{T}{2E_{F1}}\right)}^{\frac{\omega}{v_{{2}}}}  \frac{dq\,q^3}{(v_{F1} q)^2-(\frac{q^2}{2m})^2 -\omega^2} \notag\\
    & - \int_{\frac{\omega}{v_{{2}}}}^{\left (2p_{F1}-\frac{\omega}{v_{F1}} \right)\left(1 - \frac{T}{2E_{F1}}\right)} \frac{dq\,q^3}{(v_{F1} q)^2-(\frac{q^2}{2m})^2-\omega^2} \notag\\
    & = 2m^2\biggl\{\log{\left(\frac{T}{E_{F1}}\right)}- 2\log{\left( 1- \left( \frac{\omega}{T_C}
    \right)^2 \right)}\notag \\
    &+\log{\biggl( 1+\frac{\omega}{2T}\biggr) }\biggr\};
\end{align}
and
\begin{align}
      I_4&=\int_{\frac{\omega}{v_{{2}}}}^{\left (2p_{F1}-\frac{\omega}{v_{F1}}\right )\left(1  -\frac{T}{2E_{F1}}\right)}  \frac{dq\,q^5 {v_{F2}^2}}{(v_{F1} q)^2-(\frac{q^2}{2m})^2-\omega^2}\notag\\
     & -\int_{\frac{\omega}{v_{F1}}\left(1 + \frac{T}{2E_{F1}}\right)}^{\frac{\omega}{v_{{2}}}}  \frac{dq\,q^5{v_{F2}^2}}{(v_{F1} q)^2-(\frac{q^2}{2m})^2-\omega^2} \notag\\
      &=-2m^2T_C^2 \biggl\{\left(1- \left( \frac{\omega}{T_C} \right)^2\right)- 2\log{\biggl(1- \left(\frac{\omega}{T_C}\right)^2}\biggr)\notag \\
        &+\log{\biggl(1- \frac{\omega} {2T}}\biggr)+\log{\left(\frac{T}{E_{F1}}\right)}\biggr\}.
\end{align}
We separately study the case $T<T_C$ and the case $T>T_C$.
In the first case the integration is bounded by $T$ 
we find
\begin{equation}
    \frac{1}{\tau_D}\simeq 4 {\left(\frac{J_K^2}{N} \right)^2}\frac{(C_K \nu_{1})^2}{E_{F1}} T_C^2 T^2
    \log{\left(\frac{E_{F1}}{T}\right)}
    \label{Kitaev32}
\end{equation}
In the {opposite} case $T>T_C$, the $\omega$ integral splits into two 
\begin{equation}
    \frac{1}{\tau_D}=\frac{1}{\tau_D^{(1)}}+\frac{1}{\tau_D^{(2)}}
    \label{Kitaev33}
\end{equation}
where we can define 
\begin{equation}
    \frac{1}{\tau_D^{(1)}}={\left(\frac{J_K^2}{N} \right)^2}\int_0^{T_C\left(1-\frac{T}{2E_{F1}} \right)} \frac{d\omega}{\sinh^2{\bigl( \frac{\omega}{2T} \bigr)}}I_q^C(\omega),
    \label{Kitaev34}
\end{equation}
where $I_q^C(\omega)$ defined as in \eqref{Kitaev26}, while 
\begin{equation} 
    \frac{1}{\tau_D^{(2)}}= {\left(\frac{J_K^2}{N} \right)^2}\int_{T_C}^T \frac{d\omega}{\sinh^2{\bigl( \frac{\omega}{2T} \bigr)}} I_q^{NC}(\omega),
    \label{Kitaev35}
\end{equation}
with $I_q^{NC}(\omega)$ defined as in \eqref{Kitaev22} but with a minus sign in front. We find that
\begin{equation}
  \frac{1}{\tau_D^{(1)}}\simeq{\left(\frac{J_K^2}{N} \right)^2} \frac{16}{3} \frac{(C_K \nu_{1} )^2}{E_{F1}} (T_C)^3 T
    \log{\left(\frac{E_{F1}}{T}\right)},
    \label{eq:1/tau^1Kitaev}
\end{equation}  
and 
\begin{equation}
     \frac{1}{\tau_D^{(2)}}\simeq 4 {\left(\frac{J_K^2}{N} \right)^2}\frac{(C_K \nu_{1} )^2}{E_{F1}} T^4
    \log{\left(\frac{E_{F1}}{T}\right)}.
\end{equation}
In the $T/T_C \gg 1$ limit $\frac{1}{\tau_D^{(1)}} {\ll} \frac{1}{\tau_D^{(2)}}$, therefore
$\frac{1}{\tau_D}\simeq {\frac{1}{\tau_D^{(2)}}}.$
Our results are summarized in Eq.~\eqref{eq:tauDKitaevC} and \eqref{eq:tau0KitaevC} and shown in panels c) and d) of Fig.~\ref{fig:Kitaev2Regimes}.

\subsection{QSL with Fermi surface ($\mathbb{Z}_2$ and U(1))}

As in the case of the Kitaev QSL, we distinguish the situation of slow and fast spinons. For $v_{F2}>v_{F1}$ both the real and the imaginary parts of the QSL polarization operator contribute, while for
$v_{F2}<v_{F1}$ there is a 
region $0<q<\frac{\omega}{v_{F2}}\left(1+\frac{T}{2E_{F1}}\right)$, in which only the real part contributes, and $\frac{\omega}{v_{F2}}\left(1+\frac{T}{2E_{F1}}\right)<q<-\frac{\omega}{v_{F1}}\left(1-\frac{T}{2E_{F1}}\right)+2p_{F1}\left(1-\frac{T}{2E_{F1}}\right)$, in which both real and imaginary parts contribute to the QSL polarization operator, see Fig.~\ref{fig:Z2Detailed}.

\subsubsection{Limit of fast spinons}

For $v_{F2} > v_{F1}$, Eq.~\eqref{eq:ALDrag} becomes
\begin{equation}
    \begin{split}
    \frac{1}{\tau_D}=&
    {\left(\frac{J_K^2}{N}\right)^2}\frac{1}{ p_{F}^2 \,m T}\int_{0}^{\infty}
    d\omega \frac{\omega^2}{\sinh^2{\bigl(\frac{\omega}{2T} \bigr)}}\times
    \\
    & \times\int{dq}q^3 \frac{(\nu_{2} {v_{F2}}q)^2}{(v_{F2} q)^2-\omega^2} \frac{\nu^2_{1}}{(v_{F1} q)^2-\bigl( \omega^2+\bigl(\frac{q^2}{
    2m}\bigr)^2 \bigr)},\\
    \end{split}
    \label{A11}
\end{equation}
which is only valid if $p_{F2}\gg p_{F1}$.
Paying attention to the integration limits explicitly expressed in \eqref{eq:condition_int_boundary2}, and approximating 
$\omega^2\sinh^{-2}{\bigl(\frac{\omega}{2T} \bigr)}\simeq
 (2T)^2 \theta(T-\omega)$, 
{this leads to} the relaxation time reported in the main text in Eq.~\eqref{eq:tauDZ2fast}.

\subsubsection{Limit of slow spinons}

Now we turn to the case $v_{F2}<v_{F1}$ for which
Eq.~\eqref{eq:ALDrag} can then be written as
\begin{equation}
    \frac{1}{\tau_D}={\left(\frac{J_K^2}{N}\right)^2}\frac{(2T)^2}{ p_{F1}^2 m T}\int_0^{T}d\omega  \{I_q^{(A)}+ {I_q^{(B)}}\},
    \label{A14}
\end{equation}
with
\begin{equation}
    I_q^{(A)}=\int_{\frac{\omega}{v_{F1}}\left(1+\frac{T}{2E_{F1}}\right)}^{\frac{\omega}{v_{F2}}\left(1+\frac{T}{2E_{F1}}\right)} \frac{dq q^3(\nu_{1} \nu_{2})^2}{(v_{F1} q)^2-(\omega^2+\bigl( \frac{q^2}{2m} \bigr)^2  )};
\end{equation}
\begin{equation}
\begin{split}
    I_q^{(B)}=&\int_{\frac{\omega}{v_{F2}}\left(1+\frac{T}{2E_{F1}}\right)}^{{\left (2p_{F1}-\frac{\omega}{v_{F1}} \right )}\left(1-\frac{T}{2E_{F1}}\right)} \frac{dq q^3(\nu_{2} {v_{F2}}q)^2}{(v_{F2} q)^2-\omega^2}
    \\
    &\times \frac{\nu^2_{1} }{(v_{F1} q)^2-\bigl( \omega^2+\bigl(\frac{q^2}{2m}\bigr)^2 \bigr)}.
\end{split}
\label{}
\end{equation}
In the presence of a Fermi surface, the crossover temperature is
$T_C=2v_{F2}p_{F1}$. 
We find that
\begin{equation}
    \begin{split}
    &I_q^{(A)}+I_q^{(B)}=
    2(\nu_{2} \nu_{1} m)^2 \biggl\{ \log{\left(1- \left( \frac{\omega}{T_C}\right)^2\right)} +\\
    &+  \frac{1}{16}\frac{1}{1-\bigl(
    \frac{\omega}{T_C}\bigr)^2}
    \biggl[ \log{\biggl( \frac{E_{F1}}{T}\biggr)}-\log{\biggl(1+\frac{\omega}{2T}\biggr)}
    +\\
    & +\log{\biggl(1-\biggl(\frac{\omega}{T_C}\biggr)^2\biggr)}\biggr]
    \biggr\}.
    \end{split}
    \label{A17}
\end{equation}
Therefore we found that the integral in \eqref{A14} has the form of Eq.~\eqref{eq:tauDZ2slow1} for $T<\omega_C$. We can argue that this is consistent with a smooth crossover from $v_{F2}>v_{F1}$ to $v_{F2}<v_{F1}$.
In the case of $T>T_C$ the omega integral again splits into two, allowing the definition of the scattering rate as
\begin{equation}
    \frac{1}{\tau_D}=\frac{1}{\tau_D^{(1)}}+\frac{1}{\tau_D^{(2)}},
    \label{eq:tauSeparation}
\end{equation}
where we can define 
\begin{equation}
    \frac{1}{\tau_D^{(1)}}={\left(\frac{J_K^2}{N}\right)^2\frac{1}{ p_{F1}^2 m T}}\int_0^{T_C} \frac{d\omega \, {\omega^2}}{\sinh^2{\bigl( \frac{\omega}{2T} \bigr)}}I_q^C(\omega),
    \label{A20}
\end{equation}
where $I_{q}^C(\omega)=I_{q}^{(A)}+I_{q}^{(B)}$, and 
\begin{equation}
    \frac{1}{\tau_D^{(2)}}= {\left(\frac{J_K^2}{N}\right)^2\frac{1}{ p_{F1}^2 m T}}\int_{T_C}^T \frac{d\omega
 \, {\omega^2}}{\sinh^2{\bigl( \frac{\omega}{2T} \bigr)}} I_q^{NC}(\omega),
    \label{A21}
\end{equation}
with
\begin{equation}
    I_q^{NC}=\int_{\frac{\omega}{v_{F1}}\left(1+\frac{T}{2E_{F1}}\right)}^{-\frac{\omega}{v_{F1}}+2p_{F1}\left(1-\frac{T}{2E_{F1}}\right)} dq  \frac{(\nu_{1} \nu_{2})^2q^3}{(v_{F1} q)^2-(\frac{q^2}{2m})^2- \omega^2}.
\end{equation}
We thus find
\begin{equation}
    \frac{1}{\tau_D^{(1)}}\simeq{\left(\frac{J_K^2}{N}\right)^2}\frac{(\nu_{2} \nu_{1})^2}{ E_{F1}} \frac{T_C}{T}T^2\,\biggl(6\log{\left(\frac{E_{F1}}{T}\right)-340\biggr)},
    \label{A23}
\end{equation}
and
\begin{equation}
    \frac{1}{\tau_D^{(2)}}\simeq \frac{(2\nu_{2}\nu_{1})^2}{ E_{F1}}T^2\log{\left( \frac{E_{F1}}{T} \right)}.
    \label{A24}
\end{equation}
Because of the suppressing term $(T_C/T)$, we conclude that
$\frac{1}{\tau_D}\simeq \frac{1}{\tau_D^{(2)}}$, as declared in Eq.~\eqref{eq:tauDZ2slow2}.

\section{Theory with momentum relaxation within QSL layer}

In this Appendix we present details on the calculation of the drag resistivity in the case when momentum is relaxed within the QSL layer. We focus on the situation of a spinon Fermi surface.

\subsection{{Collision integral from} {Metal-QSL coupling}}
\label{app:KondoDerivation}

In this section, we present details on the derivation of the collision integral accounting for the interlayer momentum transfer, see Sec.~\ref{sec:MomentumTransfer} of the main text. This calculation is based on the Kondo interaction Hamiltonian describing the coupling between spinons and the electrons in the metal, Eq.~\eqref{eq:KondoCouplingLOCAL}. This Hamiltonian gives rise to a term in the self energy of the spinon $\hat{\Sigma}(p)$ ($p=(\v {p},\epsilon)$) as displayed in Fig.~\ref{fig:sunsetdiagram} and enters the Keldysh component of the Dyson equation leading to~\cite{Kamenev2011}
\begin{equation}
    -i \left(\big[G^R_i\big]^{-1}\circ F_i - F_i \circ \big[G^A_i\big]^{-1}\right) = \mathbb{I}^{coll}_i
\end{equation}
with `$\circ$' denoting concatenation of operators and
"off-shell" collision integral
\begin{equation}
    \mathbb{I}^{coll}_{{i}}(p)=i\left[\Sigma^K_{{i}(p)}-(\Sigma^R_{{i}}(p)-\Sigma^A_{{i}}(p))F_{K{i}}(p)\right].
    \label{eq:OffshellInt}
\end{equation}
{H}ere the subscript ${{i}}$ 
denotes that this collision integral will enter the Boltzmann equation of layer $i$ ({$i = 2$ for spinons}), the superscripts 
$K$, $R$, and $A$ stand for the Keldysh, retarded, and advanced components of the spinon self energy, respectively, and $F_i(\v {p},\omega)$ is related to the (quasi-) distribution function (see Eq.~\eqref{on-shell-dfunc} below).

The self energy components {are} 
\begin{align}
    \label{SelfEnegy_Keldysh}
    \Sigma^K_i(p)&=\left(\frac{J_K^2}{2N}\right) \int dq\,dk\,dk' G_i^{R-A}(q)G^{R-A}_j(k)G^{R-A}_j(k') \notag\\
    &\times \{F_{i}(q)
    (F_j(k)F_j(k')-1)-(F_j(k)-F_j(k'))\notag\\
    &\times \delta(p+k'-q-k),
\end{align}
where $G^{R-A} = G^R- G^A$ is the difference of retarded and advanced Green's functions and
\begin{equation}
\label{SelfEnegy_(R-A)F}
    \begin{split}
    &(\Sigma^{R-A})_i(p)F_{i}(p)= \left(\frac{J_K^2}{2N}\right) \int dq dk dk' G^{R-A}_i(q)G^{R-A}_j(k)\times \\&\times G^{R-A}_j(k') \{F_{i}(p) (F_j(k)F_j(k')-1) +\\
    &+ F_i(p)F_j(q)(F_j(k')-F_j(k))\}\times \delta(p+k'-q-k).
    \end{split}
\end{equation}
Here, ${G_i}(\v {k},\omega)$ is the dressed Green's function in layer $i$.
The integration is defined as $\int dp ={\frac{1}{2\pi}}\int (dp)d\omega$.
 
The on-shell distribution function is defined as
\begin{equation}
    2f_i-1  = \begin{cases} \int\frac{d{\xi}}{2\pi} ({G}^R_i-{G}^A_i)(p)F_i(p), & {i = 2 \text{ and U(1) QSL},}\\ {\int\frac{d\epsilon}{2\pi} ({G}^R_i-{G}^A_i)(p)F_i(p),} & {\text{else.}}\end{cases} \label{on-shell-dfunc}
\end{equation}
Here, $f_2 = f_2 (\epsilon, \hat p)$ is implied for layer 2 in the case of the U(1) QSL and $f_i = f_i(\v  p)$ in all other cases.
To derive the "on-shell" collision integral ($I^{coll}_i$), we need to multiply by $\frac{G^{R-A}_i}{{2\pi}}$ and perform an additional external integration over $\xi$ (in the case $i = 2$ and a U(1) QSL) or $\epsilon$ (in all other cases):
\begin{equation}
    I^{coll}_{i \rightarrow j}= \begin{cases} \int \frac{d\xi}{{2\pi}} (G^R_i-G^A_i)(p)\mathbb{I}^{coll}_i(p), & {i = 2 \text{and U(1) QSL},} \\
    \int \frac{d\epsilon}{{2\pi}} (G^R_i-G^A_i)(p)\mathbb{I}^{coll}_i(p), & \text{{else}.}
    \end{cases}
    \label{on-shell Coll Integral}
\end{equation}
Finally, using Eqs.~\eqref{eq:OffshellInt}, \eqref{SelfEnegy_Keldysh} to \eqref{on-shell Coll Integral} we obtain the collision integral 
\begin{widetext}
\begin{equation}
    \begin{split}
    I^{\rm{coll}}_{j \rightarrow i}(\v {p}_i)=& (2\pi)^3 \frac{J_K^2}{N}\int (d {p}_{i'}) (d{p}_j) (d{p}_{j'}) \delta(\v {p}_i+\v {p}_{j}-\v {p}_{i'}-\v {p}_{j'})\delta(\epsilon_i+\epsilon_j-\epsilon_{i'}-\epsilon_{j'})\times\\
    \times & \left\{ f(\v {p}_{i'})f(\v {p}_{j'})[1-f(\v {p}_{i})][1-f(\v {p}_{j})]- f(\v {p}_{i})f(\v {p}_{j})[1-f(\v {p}_{i'})][1-f(\v {p}_{j'})] \right\}.
    \end{split}
    \label{eq:metal-qsl}
\end{equation}
\end{widetext}
This collision integral describes the momentum and energy transferred from layer $j$ to $i$ due to Kondo coupling. We stress that it is valid both for a $\mathbb{Z}_2$ and a U(1) QSL {assuming that the measure of integration is interpreted accordingly, see Eqs.~\eqref{def:misura1}--\eqref{def:misura2}}. 
Using the linear expansion about the equilibrium distribution function, Eq.~\eqref{eq:deltafdefinition}, this concludes the derivation of Eq.~\eqref{eq:metal-qsl-linearized} of the main text. A 
factor of $(2\pi)^3$ is reabsorbed into $J_K$ in the rest of the paper concerning the Boltzmann approach.

\subsection{General solution for drag rate}
\label{app:GeneralFormula}

We proceed to solve the system of equations \eqref{eq:system} 
to calculate the drag current defined in Eq.~\eqref{eq:DefDragCurrent}.
We remark that to the leading order in $J_K$
\begin{equation}
    \psi_1 \sim O(J_K^0) \longrightarrow \psi_2 \sim O(J_K^2) \longrightarrow  \psi_3 \sim O(J_K^4).
    \label{remark}
\end{equation}
As the first step, we plug Eq.~\eqref{B1 c} into \eqref{eq:DefDragCurrent} 
and obtain
\begin{equation}
    \v {j}_3=-\frac{e \, \tau_3 }{m_3}\int (d{p}_3) I^{\rm{coll}}_{2\rightarrow3}\,\v {p}_3.
\end{equation}
Using Eq.~\eqref{eq:metal-qsl-linearized} and the solution of Eq.~\eqref{B1 1},
\begin{equation}
    \psi_1= -e\frac{1}{m_1 T}\tau_1 \v {E}_1\cdot \v {p}_1,
    \label{B6}
\end{equation}
the drag current can be written as  
\begin{align}
    \v {j}_3&= -\frac{e\,\tau_3}{2m_3}\frac{J_K^2}{N}\int
    (d{p}_{2})(d{p}_{2'})(d{p}_{3})(d{p}_{3'})
    (\v {p}_3-\v {p}_{3'})\nonumber \\ &\times\delta^{(3)}(P_2+P_{3}-P_{2'}-P_{3'}) f_2 f_3 \tilde{f}_{2'} \tilde{f}_{3'}(\psi_2-\psi_{2'})
    \label{B9}
\end{align}
where we have used the symmetry properties of the collision integral to write the term $\v {p}_3-\v {p}_{3'}$.

At the next wtep, we determine $\psi_2$ from Eq. \eqref{B1 b}. 
The sum of the two collision integrals can be written as
\begin{equation}
     \begin{split}
     &f_2 \tilde{f}_2 \psi_2= \tau_2(I^{\rm{coll}}_{1{\rightarrow}2}+I^{\rm{coll}}_{3\rightarrow2})\simeq\\
     & \simeq \frac{J_K^2 \tau_2}{N}\int
    (d{p}_{2''})(d{p}_{1})(d{p}_{1'})
    f_2 f_1 \tilde{f}_{2''} \tilde{f}_{1'}(\psi_1-\psi_{1'})\\
    & \times \delta^{(3)}(P_2+P_{1}-P_{2''}-P_{1'}),
    \end{split}
    \label{B10}
\end{equation}
in which we have neglected the higher orders in $J_K$. 
We can use the symmetry properties of $\v {j}_3$ and write
\begin{equation}
\begin{split}
    & \v {j}_3=  -\frac{e\,\tau_3}{m_3}\frac{J_K^2}{N}\int
    (d{p}_{2})(d{p}_{2'})(d{p}_{3})(d{p}_{3'})
    (\v {p}_3-\v {p}_{3'})\times\\
    & \times \delta^{(3)}(P_2+P_{3}-P_{2'}-P_{3'}) f_2 f_3 \tilde{f}_{2'} \tilde{f}_{3'}\psi_2.
    \end{split}
 \label{B9}
\end{equation}
Subsequently we use Eq. \eqref{B10} and the relations
\begin{equation}
\begin{split}
&\delta^{(3)}(P_2+P_{3}-P_{2'}-P_{3'})=\int dQ (2\pi)^3 \delta^{(3)}({P_3-P_{3'}-Q}),\times \\
&\times \delta^{(3)}({P_2-P_{2'}+Q})  \\
&\delta^{(3)}(P_2+P_{1}-P_{2''}-P_{1'})=\int dQ' (2\pi)^3 \delta^{(3)}({P_2-P_{2''}-Q'})\times \\ &\times \delta^{(3)}({P_1-P_{1'}+Q'}), \\
\end{split}
\end{equation}
to rewrite the current in the following way:
\begin{equation}
\begin{split}
    & \v {j}_3=\lambda \int dQ\,dQ' \int d\{({p})\}(\v {p}_3-\v {p}_{3'})(\v {p}_{1}-\v {p}_{1'})\cdot \v {E}_1\times\\
    &\times \delta^{(3)}({P_3-P_{3'}-Q}) \delta^{(3)}({P_{2''}-P_{2'}+Q})\times\\
    &\times \delta^{(3)}({P_{2''}-P_{2}-Q'})\delta^{(3)}({P_{1'}-P_{1}+Q'}) \times\\
    &(f_3 \tilde{f}_{3'}) (f_{1} \tilde{f}_{1'}) (f_{2} \tilde{f}_{2'}), \\
\end{split}
\end{equation}
where the coefficient is defined as $\lambda=-e^2 \frac{\tau_1 \tau_2 \tau_3}{m_1 m_3 T}(2\pi)^{10} \left(\frac{J_K^2}{N}\right)^2$
and we used the shorthand notation for the integration measure 
$d\{({p})\}$. We make use of the identity
\begin{equation}
    f_p\tilde{f}_{p'}=\frac{f_p-f_{p'}}{1-e^{(\epsilon_p - \epsilon_{p'})/T}}
    \label{B13}
\end{equation}
and the energy conserving delta functions to get 
\begin{equation}
\begin{split}
    &\v {j}_3=-e^2\frac{\tau_1\tau_2\tau_3}{2m_1m_3T}\left(\frac{J_K^2}{N}\right)^2(2\pi)^4 \times \\
    &\times \int dQdQ'\frac{{\mathrm{Im}}[\Pi_3^{R}(Q)] {{\mathrm{Im}}}[\Pi_1^{R}(Q')] I_2(Q,Q')}{\sinh{\bigl( \frac{\omega}{2T} \bigr)} \sinh{\bigl( \frac{\omega'}{2T} \bigr)} \sinh{\bigl( \frac{\omega+\omega'}{2T} \bigr)}}\v {q} (\v {q}'\cdot \v {E}_1),\\
\end{split}
\label{eq:Current-GeneralFormula}
\end{equation}
where we have used the definition Eq.~\eqref{eq:I2DefMainText} of the main text. Expression in Eq. \eqref{eq:Current-GeneralFormula} is  
valid both for the $\mathbb{Z}_2$ and the U(1) QSL {and the origin of Eq.~\eqref{eq:DragGeneralMainText} of the main text}.

\subsection{$\mathbb{Z}_2$ QSL: Fermi liquid behavior}

Here we calculate $I_2(Q,Q')$ for the $\mathbb{Z}_2$ QSL. We take into account also the integration over $\hat{q}$ and $\hat{q}'$, so evaluating the object
\begin{equation}
    \begin{split}
    &\langle I_2(Q,Q')(\vec{q} \cdot \hat{E}_1)(\vec{q}' \cdot \hat{E}_1) \rangle_{\hat{q},\hat{q}'}= \int \frac{d^2p_{2}''}{(2\pi)^2}\\
    &\bigl\langle
    (\vec{q} \cdot \hat{E}_1)\delta(\omega_q^{\pm}-\v {v}_2^{''}\cdot \v {q})\rangle_{\hat{q}}\,\, \langle(\vec{q} \cdot \hat{E}_1) \delta(\omega'^{-}_{q'}-\v {v}_2^{''}\cdot \v {q}')\rangle_{q'} \times \\
    &\times (f(\epsilon_{p_2''}-\omega')-f(\epsilon_{p_2''}+\omega))
    \bigr\rangle_{\hat{q},\hat{q}'},
    \end{split}
    \label{BZ219}
\end{equation}
where we have defined $\omega_q^{\pm}= \omega \pm \frac{q^2}{2m_2}$.
We calculate the object 
\begin{equation}
    \langle (\v{q} \cdot \hat{E}_1)\delta(\v {v}_2'' \cdot \v {q}-\omega_q^{\pm}) 
    \rangle_{\hat{q}}= \frac{\hat{p}_2''\cdot \hat{E}_1}{\pi v_{{2}}''} 
    \frac{\omega_q^{\pm}}{\sqrt{(v_{{2}}''q)^2-\omega_q^{\pm 2}}},
\end{equation}
and use the approximation 
\begin{equation}
    f(\epsilon_{p_2''}-\omega')-f(\epsilon_{p_2''}+\omega)\!\!\underbrace{\simeq}_{\omega,\omega' \ll T} \!\!2T \delta\left(\epsilon + \frac{\omega-\omega'}{2}\right)\sinh\left(\frac{\omega +\omega'}{2T}\right).
    \label{BZ222}
\end{equation}
Recalling the operative expression for the polarization operator \eqref{A8} and that $\omega\ll E_{F1}$ we can write
\begin{equation}
    \begin{split}
        &\langle I_2(Q,Q'(\vec{q} \cdot \hat{E}_1)(\vec{q}' \cdot \hat{E}_1))\rangle_{\hat{q},\hat{q}'}={-}\frac{4T \sinh{\bigl(\frac{\omega+\omega'}{2T} \bigr)}}{(2\pi)^6 \omega \omega'E_{F2}}
        \\ &\times\omega^+_q\omega^+_{q'} \mathrm{Im}\Pi_2^R(\v {q},\omega)\mathrm{Im}\Pi_2^R(\v {q}',\omega').\\
    \end{split}
\end{equation}
The drag conductivity can be the concisely expressed as
\begin{equation}
    \sigma_D=e^2 \left(\frac{J_K^2}{N} \right)^2 \frac{\tau_1 \tau_2 \tau_3}{2\pi m_1 m_2 m_3 p_{F2}^2} I_{12}I_{32}
    \label{BZ224}
\end{equation}
where we have defined
\begin{equation}
    I_{i3}=\int d\v {q}d\omega \frac{q^2}{\omega \sinh{\bigl(\frac{\omega}{2T} \bigr)}}\mathrm{Im}\Pi_i^R(\v {q},\omega)\mathrm{Im}\Pi_2^R(\v {q},\omega).
    \label{BZ224}
\end{equation}
We recognize the existence of four regimes {for $I_{i2}$, $i = 1,3$)}:
$(\frac{v_{F2}}{v_{Fi}}>1,\frac{v_{F2}}{v_{Fi}}<1)\times(\frac{p_{F2}}{p_{F{i}}}>1,\frac{p_{F2}}{p_{Fi}}<1)$.
{ We start with the limit $(\frac{v_{Fi}}{ v_{F2}}>1  \land \frac{p_{Fi}}{ p_{F2}}>1) \lor (\frac{v_{Fi}}{ v_{F2}}<1  \land \frac{p_{Fi}}{ p_{F2}}<1) $.}
{In this limit the Fermi energy of either the metallic plates or the QSL bounds the energy integration as show in panel d) and f) of Fig.~\ref{fig:Z2fourRegimes}.} We find that 
\begin{equation}
    I_{i2}=\left(\frac{T}{\pi} \right)^2 \frac{\nu_i \nu_{2}}{v_{Fi} v_{F2}}4p^2_{\text{min}},
    \label{BZ226}
\end{equation}
and, using Eq.~\eqref{BZ224} we find 
\begin{equation}
    \frac{\sigma_D}{e^2}= \left(\frac{J_K^2}{N} \right)^2 \frac{2\tau_1 \tau_2 \tau_3p^4_{\text{min}}T^4}{\pi^5 m_1 m_2 m_3 p_{F2}^2}\left(\frac{\nu_{1} \nu_{2}}{v_{F1}v_{F2}}\right)^2.
    \label{BZ227}
\end{equation}
The results for the drag rate in the symmetric case 
are presented in Eq.~\eqref{eq:tauDZ2BoltzNC} and Eq. \eqref{eq:tau0Z2NC}, and shown in panels {c) and e)} of Fig.~\ref{fig:Z2fourRegimes}.

We now explore the regime in which  $(\frac{v_{Fi}}{ v_{F2}}>1  \land \frac{p_{Fi}}{ p_{F2}}<1) \lor (\frac{v_{Fi}}{ v_{F2}}<1  \land \frac{p_{Fi}}{ p_{F2}}>1) $. As shown in panels {b) and h) of}  
Fig.~\ref{fig:Z2fourRegimes}, we can define a crossover temperature $T_C=2v_{\text{min}}p_{\text{min}}$. We find that 
\begin{equation}
    I_{i2}=\frac{4\nu_i \nu_{2}p^2_{\text{min}}T^2}{\pi^2v_{Fi} v_{F2}}\mathrm{min}\left(1,\frac{T_C}{T}\right).
    \label{BZ227}
\end{equation}
Squaring the term $I_{i2}$, where we now take $v_{F1}=v_{F3},p_{F1}=p_{F3}$, and using Eq. \eqref{BZ224}, we obtain 
\begin{equation}
    \frac{\sigma_D}{e^2}= \left(\frac{J_K^2}{N} \right)^2 \frac{\tau_1 \tau_2 \tau_3}{2\pi^8 E_{F2} }\frac{T^4p_{\text{min}}^4}{v_{F1}^2 v_{F2}^2}  \mathrm{min}\left(1,\frac{T_C}{T}\right)^2,
    \label{BZ228}
\end{equation}
hence in the symmetric case we obtain the results presented in Eq.~\eqref{eq:tauDZ2BoltzC} and \eqref{eq:tau0Z2NC}, and plotted in 
panels a) and {g)} in Fig.~\ref{fig:Z2fourRegimes}.

\subsection{U(1) QSL: non-Fermi liquid behavior}

Analogously to the $\mathbb{Z}_2$ case, we need to solve Eq.~\eqref{eq:Current-GeneralFormula} for a U(1) QSL, so we first need to calculate the quantity $I_2(Q,Q')$. Using the definitions \eqref{def:misura2} we can write
\begin{equation}
    \begin{split}
    & I_2(Q,Q')=\nu_{2}^3\int d\hat{p}_{2}d\hat{p}_{2'}d\hat{p}_{2''} d\epsilon_{2}d\epsilon_{2'} d\epsilon_{2''}\\
    &\delta^{(2)}({p_{F2}\hat{p}_{2''}-p_{F2}\hat{p}_{2'}+\v {q}}) \delta^{(2)}({p_{F2}\hat{p}_{2''}-p_{F2}\hat{p}_{2}-\v {q}'})\times\\
    &\times \delta(\epsilon_{2''} -\epsilon_{2'} + \omega) \delta(\epsilon_{2''} -\epsilon_{2} - \omega')
    (f(\epsilon_2)-f(\epsilon_{2'})).\\
    \end{split}
\end{equation}
Integrating out $\delta(\epsilon_{2''} -\epsilon_{2} - \omega')\delta(\epsilon_{2''} -\epsilon_{2'} + \omega)$ and integrating in $d\epsilon_2$ we find
\begin{equation}
\begin{split}
    & I_2(Q,Q')=\nu_{2}^3 {(\omega+\omega')}\int d\hat{p}_{2}d\hat{p}_{2'}d\hat{p}_{2''}  \delta^{{(2)}}(p_{F2}\hat{p}_{2''}+\\
    &-p_{F2}\hat{p}_{2'}+\v {q})\times \delta^{{(2)}}({p_{F2}\hat{p}_{2''}-p_{F2}\hat{p}_{2}-\v {q}'}).
\end{split}
\label{B20}
\end{equation}
We use the relation $\int d\hat{p}=\frac{2\pi}{p_F}\int d\v {p}\delta(|\v {p}|-p_F)$ for the integrations over $d\hat{p}_{2'}$ and $d\hat{p}_{2''}$ to rewrite Eq. \eqref{B20} as
\begin{equation}
    \begin{split}
    &I_2(Q,Q')=\nu_2^3\left(\frac{2\pi}{p_{F2}}\right)^2 
    {(\omega+\omega')}\\ 
    & \times \int dp_2^" \delta(|p_{F2}\hat{p}_{2"}+\v {q}|-p_{F2})\delta(|p_{F2}\hat{p}_{2"}-\v {q}'|-p_{F2}) .\\
    \end{split}
    \label{B22}
\end{equation}
It can be shown that 
\begin{equation}
\begin{split}
    & {\bigl \langle I_2(Q,Q')(\v {q}\cdot\hat{E_1})(\v {q}'\cdot\hat {E}_1)\bigr\rangle_{\hat{q},\hat{q}}}=\\
    &=-\frac{\nu_2^3}{\pi p_{F2}^2}{({\omega+\omega'})}\frac{\theta(2p_{F2}-q)q}{\sqrt{(2p_{F2})^2-q^2}} \frac{\theta(2p_{F2}-q') q'}{\sqrt{(2p_{F2})^2-q'^2}}
    ,
    \end{split}
    \label{B23}
\end{equation}
so that the expression used to calculate the drag conductivity is
\begin{align}
\sigma_{D}=\varsigma\int 
\frac{(\omega+\omega')dq  dq' d\omega  d\omega'}{\sinh{\bigl( \frac{\omega}{2T} \bigr)}\sinh{\bigl( \frac{\omega'}{2T} \bigr)}\sinh{\bigl( \frac{\omega+\omega'}{2T} \bigr)}}
\nonumber \\ \times
\frac{q^2\,{{\rm{Im}}}[\Pi_3^R(q,\omega)]}{\sqrt{(2p_{F2})^2-q^2}} \frac{q'^2\,{\rm{Im}}[\Pi_1^R(q',\omega')]}{\sqrt{(2p_{F2})^2-q'^2}},
    \label{B24}
\end{align}
where the integrations over the modulus of the exchanged momenta is restricted to $0<q<2p_{F2}$ and the integration in the frequencies takes has the domain $-\infty<\omega<\infty$.
The prefactor is 
\begin{equation}
    \varsigma=\frac{e^2}{8 \pi^3}\left( \frac{J_K^2}{N}\right)^2 \frac{\nu_{2}^3}{T} \frac{\tau_1 \tau_2 \tau_3}{m_1 m_3 p_{F2}^2}
    \label{prefactor}
\end{equation}

\subsubsection{{Limit of small spinon Fermi momentum}}
For $\{p_{F1,F3}\}\gg p_{F2}$ we use Eq. \eqref{A9} for the polarization operator.
We write $\sigma_D$ as 
\begin{equation}
    \sigma_D=\varsigma C(T,p_{F1},p_{F2},p_{F3})
    \label{eq:sigmaD_simple}
\end{equation}
and we only focus on the $C(T,p_{F1},p_{F2},p_{F3})$ term. 
To {extract} 
a first analytical estimate we approximate the function of $\omega$ and $\omega'$ like 
\begin{equation}
    \begin{split}
     &\frac{\omega +\omega' }{\sinh{\bigl( \frac{\omega}{2T} \bigr)}\sinh{\bigl( \frac{\omega'}{2T} \bigr)}\sinh{\bigl( \frac{\omega +\omega'}{2T} \bigr)}}  \simeq\frac{{8T}}{\bar{\omega} \bar{\omega}'}\\ &\times\theta(T-\omega)\theta(T-\omega') \theta(T-(\omega + \omega')),
    \end{split}
    \label{B26}
\end{equation}
where $\bar{\omega}=\frac{\omega}{T}$. From Eqs.~\eqref{B24}, \eqref{B26} we obtain
\begin{equation}
    \begin{split}  
        &C(T,p_{F1},p_{F2},p_{F3})=32 T^3\int_0^{1} d\bar{\omega}\int_0^{1-\bar{\omega}}d\bar{\omega}' \frac{1}{\bar{\omega}\bar{\omega}'}\\
        &\times \iint\limits_0^{2p_{F2}} dqdq' \frac{q^2\,{\rm{Im}}[\Pi_3^R(q,\omega)]}{\sqrt{(2p_{F2})^2-q^2}} \frac{q'^2\,{\rm{Im}}[\Pi_1^R(q',\omega')]}{\sqrt{(2p_{F2})^2-q'^2}}.\\
    \end{split}
    \label{B27}
\end{equation}
With little manipulation one can write
\begin{equation}
    \begin{split}
        &C(T,T_{C1},T_{C3})=32T^5\nu_1 \nu_3 (2p_{F2})^4 \int_0^{1} d\bar{\omega}\int_0^{1-\bar{\omega}}d\bar{\omega}' \\ &  
         \int_0^{1} d\bar{q} \int_0^{1} d\bar{q}' \frac{\bar{q}^2}{\sqrt{1-\bar{q}^2}} \frac{\bar{q}'^2}{\sqrt{1-{q}'^2}}\\
        &\times\frac{\theta(\frac{T_{C3}}{T}\bar{q}-\bar{\omega})}{\sqrt{(\frac{T_{C3}}{T})^2\bar{q}^2-\bar{\omega}^2}} \frac{\theta(\frac{T_{C1}}{T}\bar{q}-\bar{\omega}')}{\sqrt{(\frac{T_{C1}}{T})^2\bar{q}'^2-\bar{\omega}'^2}},
    \end{split}
    \label{B27}
\end{equation}
with $T_{Ci}=2v_{F2}p_i$.
Integrating in $q$ and $q'$ we find
\begin{equation}
    \begin{split}
     &C(T,T_{C1},T_{C3})=\frac{32 T^5}{T_{C1}T_{C3}}(2p_{F2})^4 \nu_1 \nu_3 \int_0^1 d\bar{\omega}\int_0^{1-\bar{\omega}}d\bar{\omega}' \\
     & 
     \times \theta\left[1-\bar{\omega}\left(\frac{T}{T_{C3}}\right)\right] \theta\left[1-\bar{\omega}'\left(\frac{T}{T_{C1}}\right)\right] \times \\
     &\times E\left[1-\omega \left(\frac{T}{T_{C3}}\right)^2\right] E\left[1-\omega' \left(\frac{T}{T_{C1}}\right)^2\right] ,\\
    \end{split}
    \label{B29}
\end{equation}
with E(x) being the elliptic integral of the second kind. 
There are two regimes that need to be discussed, $T\gg T_{C1,3}$ and $T\ll T_{C1,3}$, distinguished by the fact that for $T>T_{C1,3}$ the energy integral is bounded by $T_C$ and not $T$. The integration domain in the $\omega$, $\omega'$ is $(0,T_{C3})\times(0,T_{C1}) $ and the $\omega$,$\omega'$ dependence is factorized yielding 
\begin{equation}
     \begin{split}
     &C(T,T_{C1},T_{C3}) =\frac{32T^5}{T_{C1}T_{C3}}(2p_{F2})^4 \nu_1 \nu_3 \\ &\times \mathrm{min}\left(1,\frac{T_{C1}}{T} \right)\mathrm{min}\left(1,\frac{T_{C3}}{T} \right).
    \label{B30}
    \end{split}
\end{equation}
So we find that the drag relaxation rate in the non symmetric regime is
\begin{equation}
    \begin{split}
    &\frac{1}{\tau_D^{(U(1),C)}}=\frac{8}{\pi^4}\frac{(\nu_2)^3}{E_{F1} v_{F1} v_{F3}}\left( \frac{J_K^2}{N}\right)^2 \tau_2 T^4 \\
    &\times \mathrm{min}\left( 1, \frac{T_{C1}}{T}\right)\mathrm{min}\left( 1, \frac{T_{C3}}{T}\right).
    \end{split}
    \label{eq:tau_D U(1) C non-symm}
\end{equation}
In the symmetric regime the results are the ones presented in Eqs.~\eqref{eq:tauDU(1)C}, \eqref{eq:tau0U(1)NC-C} of the main text,
{and plotted in panel c) of Fig.~\ref{Fig:U(1)Analytic}.}
\\
\subsubsection{{Limit of large spinon Fermi momentum}}

{For $p_{F2}>p_{F1}$ we write the term $C(T,p_{F1},p_{F2},p_{F3})$ defined in Eq.~\eqref{eq:sigmaD_simple} as:
\begin{equation}
\begin{split}
    & C(T,p_{F1},p_{F2},p_{F3})=32 T^5 \nu_1\nu_3 \int_0^1 d\bar{\omega}\int_0^{1-\bar{\omega}}d\bar{\omega'}\\
    &\biggl(\int_{\frac{{\omega}}{2v_{3}}\left( 1+ \frac{T}{E_{F3}}\right) }^{\frac{{\omega}}{2v_{3}}\left( 1- \frac{T}{E_{F3}}\right) + 2p_{F3}\left( 1- \frac{T}{E_{F3}}\right)} dq I(q,v_{F3},m_3)\biggr)\times\\
    &\times \biggl(\int_{\frac{{\omega'}}{2v_{F1}}\left( 1+ \frac{T}{E_{F3}}\right) }^{\frac{{\omega'}}{2v_{F1}}\left( 1- \frac{T}{E_{F3}}\right) + 2p_{F1}\left( 1- \frac{T}{E_{F3}}\right)}  dq I(q,v_{F1},m_1) \biggr), 
\end{split}
\end{equation}
where the integrand of the two momenta integrals is defined as 
\begin{equation}
    I(q,v_{Fi},m_i)=\frac{ q^2}{\sqrt{(2p_{F2})^2-q^2}\sqrt{(v_{Fi}q)^2-{\omega}^2-\left( \frac{q^2}{2m_i} \right)^2}}.
\end{equation}
We calculated $C(T,p_{F1},p_{F2},p_{F3})$ to be
\begin{equation}
    C(T,p_{F1},p_{F2},p_{F3})=(32)^2\nu_1 \nu_3 m_1 m_3 \frac{E_{F3}}{T_{C3}}\frac{E_{F1}}{T_{C1}} T^5. 
\end{equation}
The drag relaxation rate in the non-symmetric case yields
\begin{equation}
    \begin{split}
    &\frac{1}{\tau_D}=\frac{128}{\pi^2} \frac{\nu_2^3 \nu_1 \nu_3 }{p_{F2}^2}
    \left(\frac{J_K^2}{N} \right)^2
    \frac{E_{F3}}{T_{C1}T_{C3}}T^4\tau_2.
    \label{eq:tauDU(1) NC Non-Symm}
    \end{split}
\end{equation}
}
{In the symmetric regime Eq.~\eqref{eq:tauDU(1) NC Non-Symm} becomes Eq.~\eqref{eq:tauDU(1)NC} of the main text, as it is plotted in panel a) of Fig.~\ref{Fig:U(1)Analytic}.}

\end{document}